\documentclass[manuscript]{aastex61}
\usepackage{natbib}
\usepackage{color}
\usepackage{enumitem}
\usepackage[caption=false]{subfig}
\usepackage{float}

\shorttitle{}
\shortauthors{Mondrik et al.}

\begin{document}
\title{An increased rate of large flares at intermediate rotation periods for mid-to-late M dwarfs}

\author{Nicholas Mondrik}
\affil{Department of Physics, Harvard University, 17 Oxford St., Cambridge, MA 02138, USA}
\affil{Harvard-Smithsonian Center for Astrophysics, 60 Garden St., Cambridge, MA 02138, USA}
\affil{LSSTC Data Science Fellow}
\email{nmondrik@g.harvard.edu}

\author{Elisabeth Newton}
\affil{NSF Astronomy \& Astrophysics Postdoctoral Fellow}
\affil{Massachusetts Institute of Technology Kavli Institute for Astrophysics and Space Research, 77 Massachusetts Avenue, Cambridge, MA 02139, USA}

\author{David Charbonneau}
\affil{Harvard-Smithsonian Center for Astrophysics, 60 Garden St., Cambridge, MA 02138, USA}

\author{Jonathan Irwin}
\affil{Harvard-Smithsonian Center for Astrophysics, 60 Garden St., Cambridge, MA 02138, USA}

\defcitealias{Newton2016}{N16}
\newcommand{\prot}{$P_\mathrm{rot}$}

\begin{abstract}
We present an analysis of flares in mid-to-late M dwarfs in the MEarth photometric survey.  We search 3985155 observations across 2226 stars, and detect 54 large ($\Delta m \geq 0.018$) flares in total, distributed across 34 stars.  We combine our flare measurements with recent activity and rotation period results from MEarth to show that there is an increase in flares per observation from low Rossby number ($R_o<0.04$, rapid rotators) to intermediate Rossby number ($0.04<R_o<0.44$, intermediate rotators) at the 99.85\% confidence level.  We additionally find an increased flare rate from the high Rossby number population ($R_o>0.44$, slow rotators) to the intermediate population at the 99.97\% level.  We posit that the rise in flare rate for intermediate $R_o$ could be due to changing magnetic field geometry on the surface of the star.
\end{abstract}

\keywords{stars: activity, stars: flare, stars: low-mass, stars: rotation}

\section{Introduction}

Stellar rotation plays an important role in many aspects of stellar physics, including informing stellar age estimates via a spin-down law \citep[eg.,][]{Skumanich1972,Barnes2007}, and contributing to false positives in radial velocity searches for exoplanets \citep[eg.,][]{Rajpaul2016}.  A large part of stellar rotation's usefulness is due to its intimate connection with the magnetic dynamo, which sustains the stellar magnetic field. The surface magnetic field is responsible for phenomena broadly defined as ``magnetic activity'', such as starspots, flares, and prominences \citep[see eg.][]{Donati2009}.

The surface magnetic fields of low-mass M dwarfs (M5-M8) appear to reside in one of two regimes: those with strong, axisymmetric dipole fields, and those with weak, higher order multipole fields \citep{Morin2010,Morin2008a}.  The Rossby number ($R_o$) of a system is commonly used to relate rotation to magnetic fields on stars.  $R_o$ is a dimensionless number defined as $\frac{P_{\mathrm{rot}}}{\tau_{\mathrm{conv}}}$, where \prot~ is the rotational period of the star and $\tau_\mathrm{conv}$ is the convective overturn timescale. $R_o$ can also be interpreted as the ratio of inertial forces to Coriolis forces (when $R_o<<1$, Coriolis forces dominate inertial forces).  Convective overturn timescales needed for calculating $R_o$ are not a direct observable; however, the ratio of X-ray luminosity to bolometric luminosity ($L_X/L_\mathrm{bol}$) is a function that can be empirically described by a flat line for \prot$\leq P_\mathrm{sat}$ (where $P_\mathrm{sat}$ is the saturation period), and by a power law given by $C (\frac{P_\mathrm{rot}}{\tau_\mathrm{conv}})^\beta$ for \prot$> P_\mathrm{sat}$. By fitting this piecewise function to stars binned by mass and varying $C$ to match solar values, the mass-dependence of $\tau_\mathrm{conv}$ can be estimated \citep[see e.g.,][]{Wright2011}.  Simulations from \citet{Gastine2013} found that there exists a bistable region of phase space below an $R_o$ of approximately 0.1 where both dipole- and multipole-dominated fields can exist in low-mass M dwarfs, based on the initial conditions of the system.   \citet{Kitchatinov2014} propose another mechanism by which two populations of magnetic field geometries might be observed; namely, magnetic field cycles that oscillate between dipole-dominated and multipole-dominated states. As an example of topological bimodality, \citet{Kochukhov2017} report that the binary system GJ65 AB, while composed of two fully convective M dwarfs with similar masses, rotational periods, and total magnetic fluxes, nevertheless show very different field geometries, with the primary being multipole dominated and the secondary showing a dipole dominated field. A major difference between dipolar- and multipolar-dominated fields is the impact of the surface magnetic field on angular momentum loss. Multipolar fields are less efficient at shedding angular momentum than dipolar fields, so it is expected that M dwarfs with weak, multipolar fields should spin down more slowly than their dipole-dominated cousins \citep{See2017}, assuming that they remain in such a state for a significant fraction of their lives. 

\citet{Newton2016} (hereafter \citetalias{Newton2016}) analyzed photometry from the MEarth Project, detecting rotation periods for 387 nearby, mid-to-late M dwarfs in the northern hemisphere.  Newton et al. (2018, submitted) extends this analysis to the southern hemisphere, providing an additional 241 rotation periods for study.  \citetalias{Newton2016} found that the rotation period distribution of mid-to-late field M dwarfs was bimodal; there exists a population of young ($\lesssim 2$ Gyr), rapidly rotating (\prot$<10$ d) stars, as well as an older (approximately 5 Gyr), slowly rotating (\prot$>70$ d) population, with relatively few stars in the intermediate region.  Moreover, \citet{Newton2017} found that rapidly rotating M dwarfs showed H-$\alpha$ in emission, while H-$\alpha$ in slow (spun-down) rotators was either not detected or seen in absorption.  Notably, the samples studied in \citet{Morin2010,Morin2008a} were all rapid rotators.  \citet{Stelzer2016a} studied early to mid M dwarfs observed as part of the K2 campaign, and found a similar dichotomy in their measures of activity between fast rotators and slow rotators, though their dividing line is placed at a period of \prot$\sim 10$ d.  This is likely due to the stellar mass difference between the K2 and MEarth surveys.  Prior to this, \citet{West2008} calibrated an age-activity relationship for M dwarfs by examining H-$\alpha$ emission and height relative to the galactic plane (a proxy for stellar age), finding that old stars were less active than young stars.

\citet{See2017} note that the results of \citetalias{Newton2016} could be explained by a change in magnetic field geometry from weak and multipolar to strong and dipolar over the lifetime of the star,  with the shift to a dipole-dominated field increasing the angular momentum loss rate and providing a physical mechanism for rapid spindown. The results of \citet{See2017} provide supporting evidence that a transition in magnetic field geometry might be responsible for the dichotomy found by \citetalias{Newton2016}. \citet{See2017} examined only stars that fall into the rapidly rotating category of \citealt{Newton2016}, and although Zeeman Dopplar Imaging (ZDI) is possible for slowly rotating stars, it yields information only about the low orders of the field's spherical harmonic expansion \citep{Donati2009}.  This would suppress an observers ability to detect highly multipolar fields on these stars.

Stellar rotation, magnetic activity, and magnetic field strength and configuration are intrinsically linked. In particular, stellar flares are caused by the reconnection of surface magnetic field loops.  Magnetic activity in late-type stars is usually measured in terms of H-$\alpha$ emission, and it is known that magnetic activity increases with decreasing stellar mass \cite[e.g.,][]{Hawley1996}.  In addition, magnetically active stars are known to flare more frequently than inactive stars.  Efforts to characterize flares on M dwarfs have been bolstered by the rise of large sky surveys such as the Sloan Digital Sky Survey (SDSS).  \citet{Kowalski2009} identified flares in SDSS Stripe 82, where the authors found that active (H-$\alpha$ in emission) M dwarfs flared at approximately 30x the rate of inactive M dwarfs.  They also found that flaring fraction increases for redder stars, which they attribute in part to a higher fraction of magnetically active stars for late-type M dwarfs.  In the context of the results from \citetalias{Newton2016}, this is compatible with lower mass stars remaining rapidly rotating (and therefore magnetically active) for longer than their more massive cousins.  Increased flare rate in late-type stars also impacts the exoplanet community, as flares release large amounts of energy in the UV portion of the stellar spectrum, which could potentially affect the atmospheres and habitability of orbiting exoplanets \citep{Segura2010}.

In addition to flare rates, flare morphologies are also of interest.  \citet{Davenport2014} study flares observed with the Kepler satellite on GJ 1243, an active M4 star, and find that a majority of flares with durations longer than approximately 30 minutes are complex in nature, meaning that the flare shows multiple peaks. \citet{Hawley2014} presents an analysis of flaring on active and inactive M dwarfs, noting that the inactive, early M dwarfs showed fewer flares than active mid M dwarfs.  Their analysis of flares on GJ 1243 revealed that the timing of flares was uncorrelated with particular starspot phase on the surface of the star, which one might expect if a specific active region on the stellar surface was responsible for generating flares.  \citet{Davenport2016} also studied flares from Kepler in the context of rotation.  The author found that the relative flare luminosity saturated at an $R_o$ of approximately 0.03, with the relative flare luminosity decreasing for increasing $R_o$ (longer periods).

The goal of this paper is to explore flares serendipitously observed with the MEarth project and to relate the presence of flares with other observables, in particular stellar rotation.  The paper is organized in the following way: section~\ref{sec:mearth_photometry} describes the MEarth Project and the photometry used in this work.  Section~\ref{sec:rotperiodrec} presents a brief overview of the extraction of rotation periods from MEarth photometry. Section~\ref{sec:flareident} details the procedure used to identify flares in MEarth data. Section~\ref{sec:results} presents the results of the flare extraction, and section~\ref{sec:conclusion} summarizes the paper.

\section{MEarth Photometry}
\label{sec:mearth_photometry}
The MEarth Project is a transit survey focused on identifying small, transiting exoplanets in the habitable zone of approximately 3000 nearby ($\lesssim33$ pc) mid-to-late M dwarfs \citep{Nutzman2007,Berta2012,Irwin2015}.  The MEarth Project is composed of two arrays of eight 40 cm telescopes, with one array in the northern hemisphere at Mt. Hopkins, AZ, and another in the southern hemisphere at Cerro Tololo Inter-American Observatory, Chile.  These installations are referred to as MEarth-North and MEarth-South, respectively.  MEarth-North has been in operation since September 2008, and MEarth-South has operated since January 2014.

In order to meet the transit detection requirements for its primary goal of detecting small planets around M dwarfs, MEarth targets are chosen to be small ($R_* < 0.33 R_\odot$), nearby stars.  \citet{Irwin2015} presents an overview of the MEarth catalog selection criteria in detail.  Because MEarth's targets are widely distributed over the sky, each target must be observed individually. The adopted MEarth cadence is approximately 20-30 minutes; this allows MEarth to monitor a large number of stars while still sampling frequently enough to detect the transit of an Earth-sized planet in the habitable zone.  While most stars are observed once every 20-30 minutes, a subset of the brightest stars are observed multiple times in a row at each pointing to average over scintillation noise due to short exposure times.  Additionally, MEarth has a trigger mode, in which a detected dip in stellar brightness can trigger additional high cadence follow-up observations, which is designed to capture a transit in real time.  These high cadence data provide a chance to sample flare light curves much more densely than typical MEarth cadence allows.

The blue end of the MEarth-North bandpass was defined in the 2008-2009 and 2009-2010 observing seasons by a 5mm thick RG715 Schott glass (longpass) filter, and the red end by the falloff of the CCD quantum efficiency.  During the 2010-2011 season, the setup was changed to a custom interference filter with a sharp red cutoff (approximating the Cousins $I$-band).  After finding that the interference filter increased systematics due to changes in humidity or precipitable water vapor rather than suppressed them, the MEarth-North bandpass was changed to a 3mm RG715 filter from 2011 onwards.  The MEarth-South array was constructed and is operated with a 3mm RG715 filter defining the bandpass.  Light curves in MEarth are not stitched together across telescopes or filters, so a single MEarth light curve is defined as a series of observations from a single telescope with a specified filter.  Looking ahead to the topic of flare identification, it is important to note that the MEarth bandpass excludes the H-$\alpha$ spectral line at 656 nm.

The MEarth reduction pipeline is based on that of \citet{Irwin2006}, with modifications described in \citet{Berta2012}. A significant systematic in MEarth noted by \citet{Irwin2011} originates from the color mismatch between (red) MEarth targets and typical (bluer) field stars.  In particular, absorption by telluric water vapor blocks more light from MEarth targets than from the bluer field stars.  Usual differential photometry techniques cannot account for this effect, and because the water vapor absorption is more dependent on temporal variability than airmass, traditional atmospheric extinction methods fail.  In order to estimate the impact of water vapor absorption, the measured differential magnitudes of all monitored M dwarfs are averaged on a half-hour timescale, producing a time series called the common mode.  This common mode is constructed separately for MEarth-North and MEarth-South.  Because the common mode is constructed from an agglomeration of monitored M dwarfs of different spectral types, it is necessary to scale the common mode before removing it from an individual M dwarf target. The scale factor is empirically fit to each individual star and is generally larger for redder stars.

The MEarth telescopes are mounted on German Equatorial Mounts, which require that the telescope assembly be flipped as the mount slews past the meridian.  This rotation of the detector plane, combined with residual flat fielding errors, serves to introduce offsets in the differential magnitude between observations taken on the eastern and western sides of the pier. Additionally, there are step functions in the photometry whenever the camera is removed for servicing. The MEarth pipeline defines observations taken on one side of the pier between camera removals as a segment.  When performing differential photometry, each segment is treated independently.

\section{Rotational Period Recovery}
\label{sec:rotperiodrec}
We present here a brief review of our period detection algorithm, described in full in \citetalias{Newton2016}. The method is based on that presented in \citet{Irwin2011} and \citet{Irwin2006}.  New rotation periods from MEarth used in this work will be published in Newton et al. (2018, submitted).

To determine a rotation period, we fit a null hypothesis and an alternative hypothesis.  The null hypothesis, $m_0(t)$, is given by

\begin{equation}
	m_0(t) = m_i + k \, c(t)
\end{equation}

where $m_i$ is the baseline magnitude for segment $i$, $k$ is the common mode scale factor, and $c(t)$ is the common mode at time $t$.  The null hypothesis model is based on a constant stellar flux at the top of the atmosphere, accounting for variable atmospheric absorption and instrument systematics. The alternative hypothesis, $m_\mathrm{alt}(t)$, is

\begin{equation}
\label{eq:alt_hypo}
	m_\mathrm{alt}(t) = m_0(t) + a_s \sin(\omega t) + a_c \cos(\omega t)
\end{equation}

where the additional terms are a linearized version of $a \sin(\omega t + \phi)$, which accounts for stellar variability due to starspots rotating into and out of view.

As noted in section~\ref{sec:mearth_photometry}, observations in MEarth are split into multiple light curves.  We fit all light curves for a given star simultaneously, and require that each star's light curves share a common period, though other parameters are allowed to float.  We perform a linear least-squares fit of Equation~(\ref{eq:alt_hypo}) on a uniformly spaced grid in frequency, corresponding to periods of 0.1 to 1500 days.  Because the null hypothesis is nested in the alternative hypothesis and we are performing a linear least-squares fit, it is appropriate to use an $F$-test to determine whether the additional sinusoidal term is statistically justified.  At each frequency in our grid, we calculate the $F$-test statistic and select the highest value as our candidate period.  The set of $F$-test values is analogous to a periodogram.

Once the best fit period is found, we examine the common mode corrected light curves by eye, to determine whether the best fit period is valid.  The guidelines we use to assess the validity of the proposed period are:

\begin{itemize}
	\item Is the period visible by eye in the phased data?
	\item Are multiple periods seen in the unphased data?
	\item Is the proposed period correlated with systematics in the model?
	\item If the candidate period is short, can variation be seen by eye in one well sampled night?
	\item If there are multiple light curves for a star, does the model fit them all?
	\item Is the period an alias of known non-astrophysical phenomenon (most commonly seen as a one day or half day period)
\end{itemize}

Upon examination, stars are given a grade of A, B, or non-rotator.  A grade rotators satisfy all of the above criteria.  B grade rotators fail in one or more of the above criteria, but are believed to be correct.  Typically, B grade rotators fail in only one criterion. Usual failure modes for B grade rotators are

\begin{itemize}
	\item Convincing detection of modulation, but lack of multiple cycles
	\item Sparse phase coverage
	\item Convincing detection of period, but amplitude is comparable to noise in the light curve.
\end{itemize}

We do not attempt to place errorbars on our periods, as there are often multiple peaks in the periodogram.  After grades are assigned, we consider A and B grade rotators on equal footing.  Looking ahead, of the 34 stars in which we detect at least 1 flare, 24 of them have a rotation period with a grade of A or B.

\section{Flare Identification}
\label{sec:flareident}
The temporal cadence and red bandpass of the MEarth project pose several challenges for detecting flares in monitored M dwarfs.  Flares are blue in color, meaning that the contrast between flaring and quiescent states is reduced in MEarth's red bandpass.  In addition, the typical cadence of MEarth (20 minutes) is not well matched with the variability timescale of most flares (of order minutes), meaning that many flare candidates have only one observation during a flare, which consists of a single bright measurement.  Despite these challenges, MEarth does have strengths as a flare survey: among them duration, precision, and scintillation--limited and trigger observing modes, the latter two of which allows the temporal sampling of MEarth to better match flare timescales.  Although most flares are short in duration, longer lasting flares are also observed and can be effectively sampled even with MEarth's nominal 20 minute cadence.  These flares, which can have decay times on the order of hours, are more likely to be complex (i.e., showing multiple peaks) in nature \citep{Davenport2014}, though MEarth lacks the temporal resolution to separate complex from classical flares.  In addition to observing longer--lived flares, MEarth's precision on a single observation is typically about $2$ mmag, meaning a robust detection of a flare candidate can be made.  We also note that we adjust our integration times based on the brightness of the target to achieve uniform precision over the survey targets; the photometric uncertainties are therefore not a function of apparent magnitude.  When operating in scintillation--limited observing mode, multiple observations taken by MEarth can be used to partially resolve the temporal evolution of even short lived flares.  Lastly, even though large or long--lived flares occur less frequently than faint or short flares, MEarth's long baseline means that there are a significant number of flares detected serendipitously despite low detection probabilities per exposure.

For pre-processing flare candidates, we make several cuts to the data.  First we require that the extinction correction be less than 0.5 mag, and also that the full-width at half maximum (FWHM) be less than seven pixels.  In addition, we restrict our analysis to observations taken with the target within 500 pixels of the center of the detector.  The first and last observation of any night are not considered for candidacy as a flare.  The positional cut discards data that was taken with bad pointings, and the FWHM and magnitude cuts discard data taken in poor conditions.

To detect flare candidates in MEarth photometry, we first correct the photometry for water vapor variation via the common mode.  To do this, we interpolate the (coarsely sampled, due to the temporal binning requirement) common mode to the time of observation, then multiply by a coefficient specific to each star and subtract from the measured photometry.  We then subtract a 1 day wide median filter centered on each observation to remove night-to-night variation.  The requirement for an observation to be classified as a flare candidate is the following:

\begin{enumerate}[label=(\roman{*}),labelsep=2em]
	\item  $m_i < 0,\, m_{i+1},\, m_{i-1}$
	\item  $|m_i| > 5\sigma_i$  
	\item  $|m_i| > 5\sigma_{\mathrm{local},i}$ 
	\item  $|m_{i+1}| > n \sigma_{\mathrm{local},i+1}$
	\item $|m_{i+1}| > n \sigma_{i+1}$
\end{enumerate}

where $m_i$ is the corrected differential magnitude of the observation as described above, $\sigma_i$ is the estimated photometric error of the observation, $\sigma_{\mathrm{local},i}$ is the standard deviation of the light curve within a 1 day box centered on the observation in question, and $n$ is a variable. When computing $\sigma_{local,i}$, the observation in question is excluded from the calculation.  When estimating $\sigma_{\mathrm{local},i+1}$, $m_{i}$ and $m_{i+1}$ are likewise excluded from the estimation. Requirement (iii) protects against outliers due to a mismatch between estimated photometric error and system performance.  For example, some rapid rotators can show significant changes in brightness over the course of one night.  This term helps reject observations due to, e.g, photometric modulations due to rotation.

{The $\sigma_\mathrm{local}$ requirement imposes limits on flare detection efficiency for rapid, high amplitude rotators.  If we approximate a rotator's light curve as $m(t) = a \sin(\omega t)$, the intrinsic standard deviation (assuming a continuous function) over one period is $\frac{a}{\sqrt{2}}$.  For our sample, the 90th quantile in rotational amplitude is approximately 13 mmag.  This would correspond to a $\sigma_\mathrm{local}$ of 9 mmag.  This approximation assumes a period of roughly 8 hours, which would place the stars amongst the most rapidly rotating in the sample.  As the period increases, the amount of phase space sampled in a night decreases, as does the intrinsic $\sigma_\mathrm{local}$.   For a period of 16 hours, which covers about $\pi$ radians of phase space in a night, the intrinsic $\sigma_\mathrm{local}$ drops to approximately $\frac{a}{3}$, which corresponds to $\sigma_\mathrm{local} \simeq 4$ mmag.  As this number is close to MEarth's per-visit precision, it does not strongly affect our ability to detect flares on most targets.}

\begin{figure}[h!]
	\centering
	\includegraphics[width=\columnwidth]{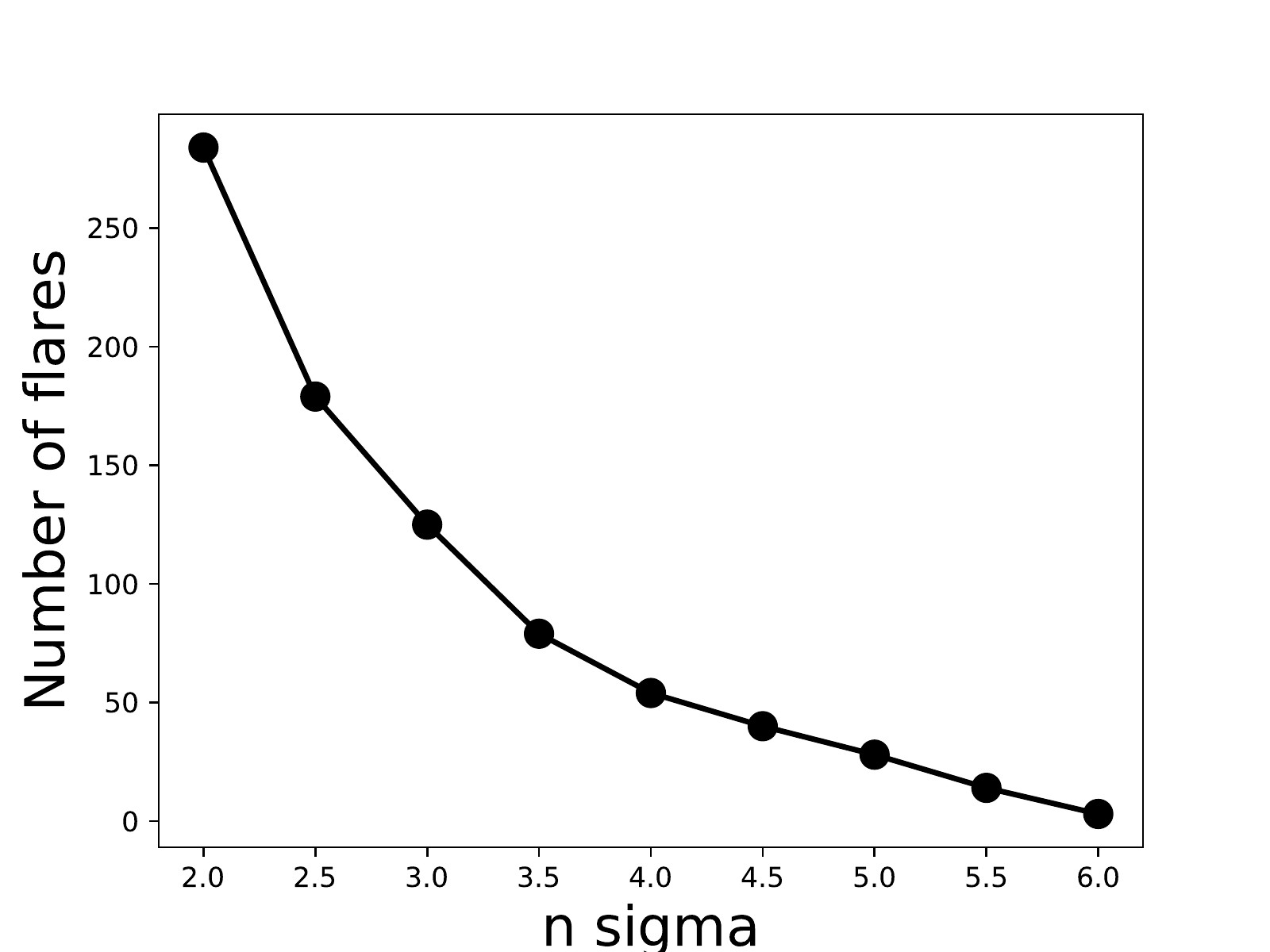}
	\caption{Curve of growth for flare candidates in the MEarth survey as a function of $n$ in criteria (iv) and (v). Linear growth ends at approximately 4$\sigma$. We therefore adopt this value for the rest of the paper.}
	\label{fig:cog}
\end{figure}

Requirements iv and v also necessitate tuning of the adjustment parameter \textit{n}. We estimate $n$ by constructing a curve of growth.  We expect that as we lower $n$, we will have more spurious detections of flares, though we will also include some true low signal-to-noise flares.  Figure~\ref{fig:cog} shows the curve of growth for a linear spacing of $n$ between 2 and 6 in steps of 0.5.  The plot is approximately linear until $n=4$, after which the number of detected flares begins to rise rapidly, presumably due to spurious events.  While some of the flares detected for the $n<4$ cases are likely real, we seek a high purity, rather than a high completeness, sample for our analysis.  We therefore adopt a value of $n=4$. We additionally do not exclude known multiple systems from the analysis.  {We also note that flares on 2MASSJ05062489+5247187 and 2MASSJ07382951+2400088 were detected simultaneously by two different telescopes in the MEarth array.  Both detections are included in the analysis below.  Since MEarth telescopes generally target stars independently of one another, leaving these flares out of the analysis would unfairly reduce flare rates.}

\section{Results}
\label{sec:results}
\subsection{Flare Characteristics}
\begin{figure*}[ht!]
	\includegraphics[width=\textwidth]{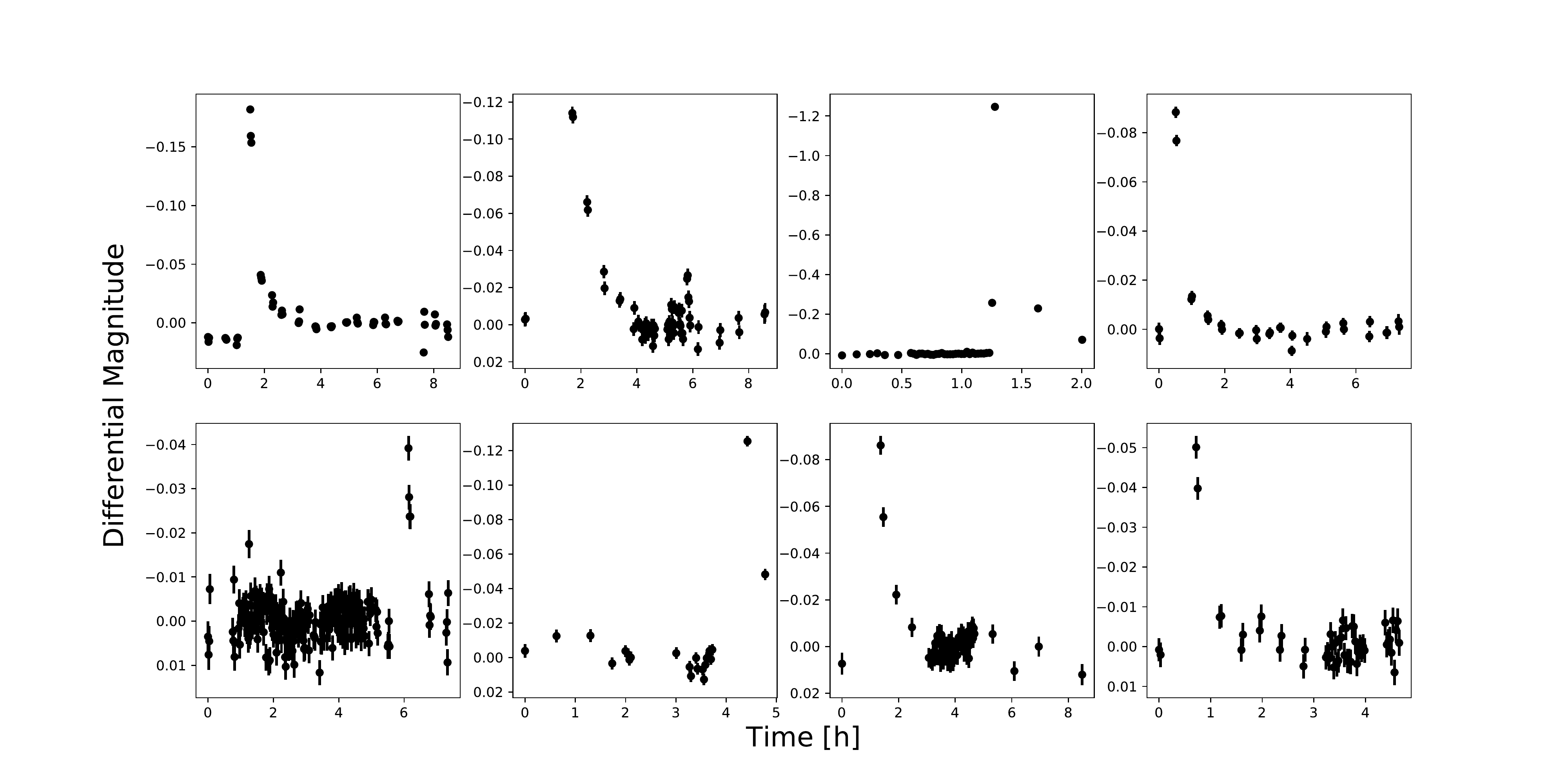}
	\caption{Examples of representative flares from the MEarth survey.  The x-axis is time in hours since the first observation of the night. For the larger flares, photometric uncertainty estimates are smaller than the points. {A figure showing all detected flares is available as an ApJ figure set.}}
	\label{fig:flares}
\end{figure*}

The above algorithm located a total of 54 events, with 34 unique stars showing at least 1 event. 24 stars showed only 1 event, 6 stars showed 2, 3 stars showed 3, and one star showed 9.  Because MEarth's detection probability for individual flares is very low, we analyze the fraction of observations that flare between different populations as defined by $R_o$ and activity (as determined by H-$\alpha$ equivalent width, EW). Some examples of flares in MEarth are shown in Figure~\ref{fig:flares}.  {Table~\ref{tab:flares} contains the minimum flare amplitudes, as well as some characteristics for flaring stars.  A negative EW in Table~\ref{tab:flares} implies emission.  We do not attempt to calculate flare energies for a number of reasons.  First, our light curves in general are too poorly sampled to empirically determine quantities such as equivalent duration, due mostly to lack of sampling during the impulsive rise phase of the flare, which is necessary to define a start time for the flares.  Using a flare template \citep[e.g., from][]{Davenport2014} to describe our observations is in principle possible; however, these large flares are very often complex \citep{Davenport2014}, meaning that they are not well-described by a template derived from a classical flare event.  In addition, flare templates derived in other bandpasses and the coefficients found therein are not guaranteed to be valid for the MEarth bandpass.  The MEarth bandpass is also not precisely known (as differential photometry requires only good stability of the system response), so integrating a spectrophotometric observation of a flaring star to measure MEarth-bandpass luminosities is not well constrained.  Any flare energies based on MEarth data would therefore be either misleading or have errorbars too large to be of much use.}

\startlongtable
\begin{longrotatetable}
\begin{deluxetable}{cccccccc}
	\tablecaption{Flaring Stars in the MEarth Dataset\label{tab:flares}}
	\tablehead{
		\colhead{2MASS Identifier} & \colhead{Flare Min. Amp. [mag]} & \colhead{T$_\mathrm{eff}$ [K]} & \colhead{Mass [$M_\odot$]} & \colhead{\prot [d]} & \colhead{EW H$\alpha$ [nm]} & 
		\colhead{\prot~Ref.} & \colhead{EW H$\alpha$ Ref.}}
	\startdata
10562886+0700527 & 0.1886 & 2946 & 0.10 & \nodata & -7.71 & \nodata & \citet{Browning2010} \\
10562886+0700527 & 0.1109 & 2946 & 0.10 & \nodata & -7.71 & \nodata & \citet{Browning2010} \\
13093495+2859065 & 0.0266 & 3269 & 0.17 &   0.21 & -6.12 & \citetalias{Newton2016} & \citet{Shkolnik2009} \\
15012274+8202252 & 0.1033 & 3246 & 0.34 & \nodata & 0.03 & \nodata & \citet{Newton2017} \\
12294530+0752379 & 0.5115 & 2861 & 0.09 & \nodata & \nodata & \nodata & \nodata \\
10564633+3246278 & 0.1532 & 3263 & 0.20 &   1.38 & -4.25 & \citetalias{Newton2016} & \citet{Newton2017} \\
06195048+6605308 & 0.0501 & 3235 & 0.39 &  20.07 & -3.48 & \citetalias{Newton2016} & \citet{Newton2017} \\
06195048+6605308 & 0.0497 & 3235 & 0.39 &  20.07 & -3.48 & \citetalias{Newton2016} & \citet{Newton2017} \\
07221747+0851566 & 0.0538 & 3209 & 0.32 &  22.90 & \nodata & \citetalias{Newton2016} & \nodata \\
07382951+2400088 & 0.0572 & 3294 & 0.36 &   3.87 & -3.70 & \citet{Hartman2011} & \citet{Shkolnik2009} \\
07382951+2400088 & 0.0363 & 3294 & 0.36 &   3.87 & -3.70 & \citet{Hartman2011} & \citet{Shkolnik2009} \\
04121693+6443560 & 0.0545 & 3275 & 0.19 &   1.59 & -3.00 & \citetalias{Newton2016} & \citet{Newton2017} \\
08215702+1748558 & 0.3358 & 3172 & 0.12 & \nodata & \nodata & \nodata & \nodata \\
05062489+5247187 & 0.1254 & 3165 & 0.13 &   0.65 & -3.91 & \citetalias{Newton2016} & \citet{Newton2017} \\
05062489+5247187 & 0.1517 & 3165 & 0.13 &   0.65 & -3.91 & \citetalias{Newton2016} & \citet{Newton2017} \\
05102012+2714032 & 0.0861 & 2836 & 0.09 &   4.61 & -49.60 & \citetalias{Newton2016} & \citet{Lepine2003} \\
01564570+3033288 & 0.0603 & 3209 & 0.32 &   1.58 & -14.95 & \citetalias{Newton2016} & \citet{Alonso-Floriano2015} \\
01564570+3033288 & 0.0744 & 3209 & 0.32 &   1.58 & -14.95 & \citetalias{Newton2016} & \citet{Alonso-Floriano2015} \\
01564570+3033288 & 0.1140 & 3209 & 0.32 &   1.58 & -14.95 & \citetalias{Newton2016} & \citet{Alonso-Floriano2015} \\
23512227+2344207 & 0.0531 & 3236 & 0.34 &   3.21 & -5.95 & \citetalias{Newton2016} & \citet{Shkolnik2009} \\
22285440-1325178 & 0.2803 & 2812 & 0.09 & \nodata & -4.40 & \nodata & \citet{Mohanty2003} \\
22480446-2422075 & 0.0500 & 3278 & 0.20 &   0.32 & -3.81 & \citet{Mamajek2013} & \citet{Bartlett2017} \\
20501615-3424424 & 0.0826 & 3246 & 0.17 & \nodata & -4.28 & \nodata & \citet{Gaidos2014} \\
20561181-1301428 & 0.1175 & 3194 & 0.15 & \nodata & \nodata & \nodata & \nodata \\
21111366-2248173 & 0.1142 & 3220 & 0.35 &  51.11 & -2.97 & \nodata & \citet{Newton2017} \\
21111366-2248173 & 0.0944 & 3220 & 0.35 &  51.11 & -2.97 & \nodata & \citet{Newton2017} \\
05532295+2212500 & 0.0712 & 3251 & 0.35 &  19.50 & -2.13 & \citetalias{Newton2016} & \citet{Newton2017} \\
05532295+2212500 & 0.0559 & 3251 & 0.35 &  19.50 & -2.13 & \citetalias{Newton2016} & \citet{Newton2017} \\
05532295+2212500 & 0.1397 & 3251 & 0.35 &  19.50 & -2.13 & \citetalias{Newton2016} & \citet{Newton2017} \\
01063072+3017112 & 0.0474 & 3153 & 0.29 & \nodata & -5.72 & \nodata & \citet{Newton2017} \\
01374851+3823318 & 0.0575 & 3232 & 0.36 &   4.10 & -6.29 & \citetalias{Newton2016} & \citet{Newton2017} \\
01374851+3823318 & 0.0436 & 3232 & 0.36 &   4.10 & -6.29 & \citetalias{Newton2016} & \citet{Newton2017} \\
01374851+3823318 & 0.1087 & 3232 & 0.36 &   4.10 & -6.29 & \citetalias{Newton2016} & \citet{Newton2017} \\
01374851+3823318 & 0.0784 & 3232 & 0.36 &   4.10 & -6.29 & \citetalias{Newton2016} & \citet{Newton2017} \\
01374851+3823318 & 0.0348 & 3232 & 0.36 &   4.10 & -6.29 & \citetalias{Newton2016} & \citet{Newton2017} \\
01374851+3823318 & 0.0181 & 3232 & 0.36 &   4.10 & -6.29 & \citetalias{Newton2016} & \citet{Newton2017} \\
01374851+3823318 & 0.0392 & 3232 & 0.36 &   4.10 & -6.29 & \citetalias{Newton2016} & \citet{Newton2017} \\
01374851+3823318 & 0.2157 & 3232 & 0.36 &   4.10 & -6.29 & \citetalias{Newton2016} & \citet{Newton2017} \\
01374851+3823318 & 0.2724 & 3232 & 0.36 &   4.10 & -6.29 & \citetalias{Newton2016} & \citet{Newton2017} \\
17462934-0842362 & 0.0814 & 3236 & 0.34 &  65.90 & -1.16 & \citet{Kiraga2012} & \citet{Gaidos2014} \\
04274130+5935167 & 0.0395 & 3260 & 0.25 &   6.85 & -2.76 & \citetalias{Newton2016} & \citet{Shkolnik2009} \\
23254016+5308056 & 0.1819 & 3213 & 0.31 &  23.54 & -4.67 & \citetalias{Newton2016} & \citet{Newton2017} \\
16553529-0823401 & 1.2462 & 2739 & 0.08 & \nodata & -7.09 & \nodata & \citet{Reiners2007} \\
12102834-1310234 & 0.0372 & 3233 & 0.36 &  42.98 & -2.24 & \nodata & \citet{Newton2017} \\
13581955-1316248 & 0.0311 & 3281 & 0.21 &   6.77 & -1.90 & \nodata & \citet{Alonso-Floriano2015} \\
14012306-0733475 & 0.0787 & 3237 & 0.36 &   0.92 & \nodata & \nodata & \nodata \\
14253413-1148515 & 0.0612 & 3370 & 0.51 &  25.01 & -1.24 & \nodata & \citet{Newton2017} \\
14294291-6240465 & 0.0884 & 3075 & 0.12 &  88.98 & -2.40 & \citet{Benedict1998} & \citet{Torres2006} \\
14294291-6240465 & 0.2495 & 3075 & 0.12 &  88.98 & -2.40 & \citet{Benedict1998} & \citet{Torres2006} \\
14294291-6240465 & 0.0803 & 3075 & 0.12 &  88.98 & -2.40 & \citet{Benedict1998} & \citet{Torres2006} \\
11575352-2349007 & 0.0363 & 3247 & 0.39 &   3.07 & \nodata & \nodata & \nodata \\
11575352-2349007 & 0.0540 & 3247 & 0.39 &   3.07 & \nodata & \nodata & \nodata \\
04575862-0506161 & 0.0884 & 3272 & 0.22 &   4.60 & \nodata & \nodata & \nodata \\
01025100-3737438 & 0.4936 & 2699 & 0.08 & \nodata & \nodata & \nodata & \nodata
	\enddata	
\end{deluxetable}
\end{longrotatetable}

\begin{figure}[h!]
	\centering
	\includegraphics[width=\columnwidth]{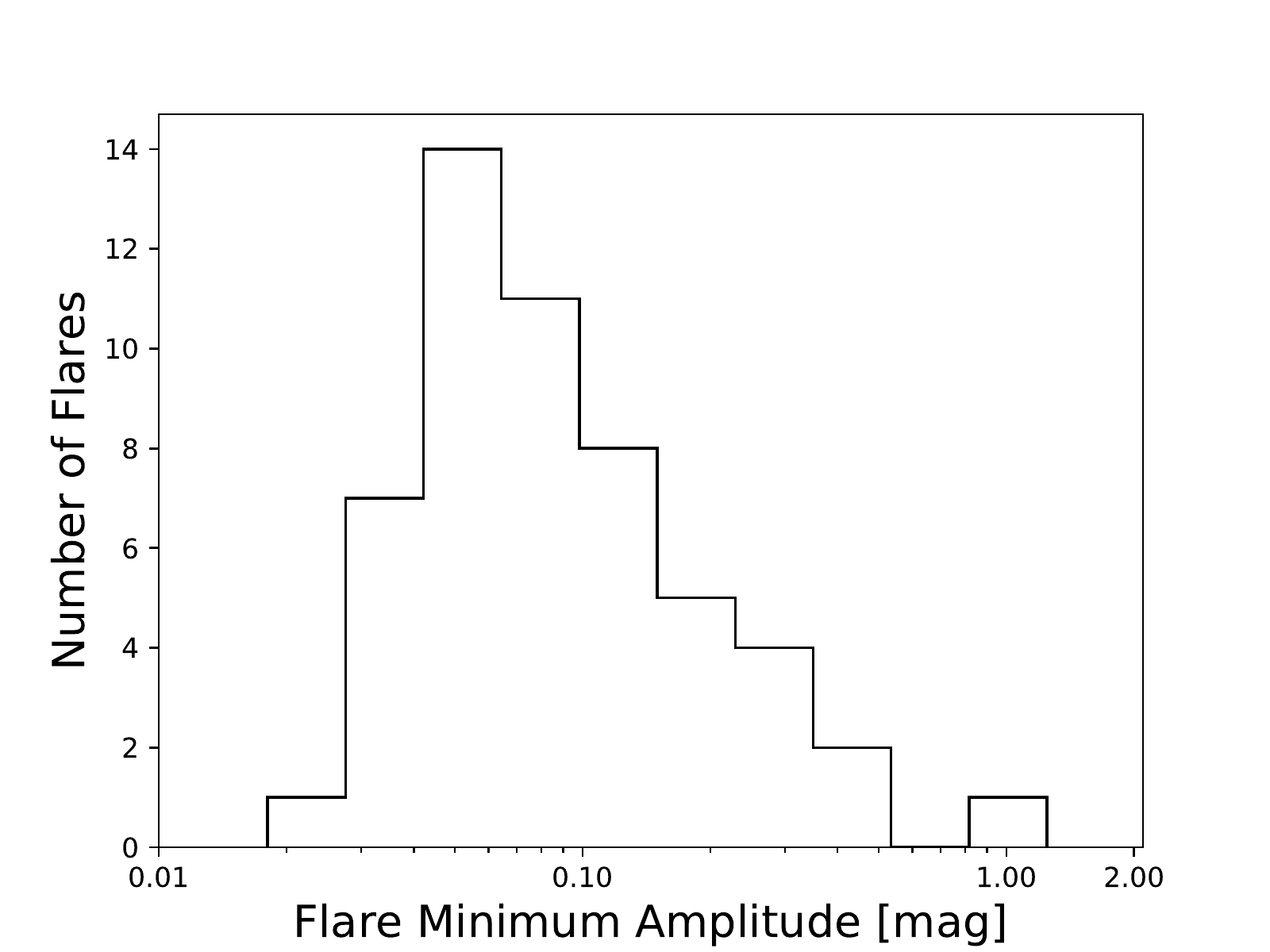}
	\caption{Amplitude distribution of flares detected in the MEarth survey.  The largest flare occurred on VB 8, a well-known flare star.}
	\label{fig:flareminamp}
\end{figure}

Because MEarth's temporal sampling is in general mismatched with the timescale for evolution of most flares, we do not attempt to calculate equivalent durations or estimate flare amplitudes.  We can, however, estimate a minimum flare amplitude by taking the brightest point in each flare lightcurve, as shown in Figure~\ref{fig:flareminamp}.  The largest flare detected in the MEarth survey had a magnitude change of 1.2 (a factor of 3 increase in brightness), and occurred on the known flare star VB 8 (0.08$M_\odot$).  The star 2MASS J01374851+3823318 (0.26$M_\odot$, \prot$=4.1$ days) flared the most frequently, with 9 flares observed in 27164 observations.

\subsection{Flares and Rotation}
\begin{figure*}
	\centering
	\subfloat[]{\includegraphics[width=0.5\textwidth]{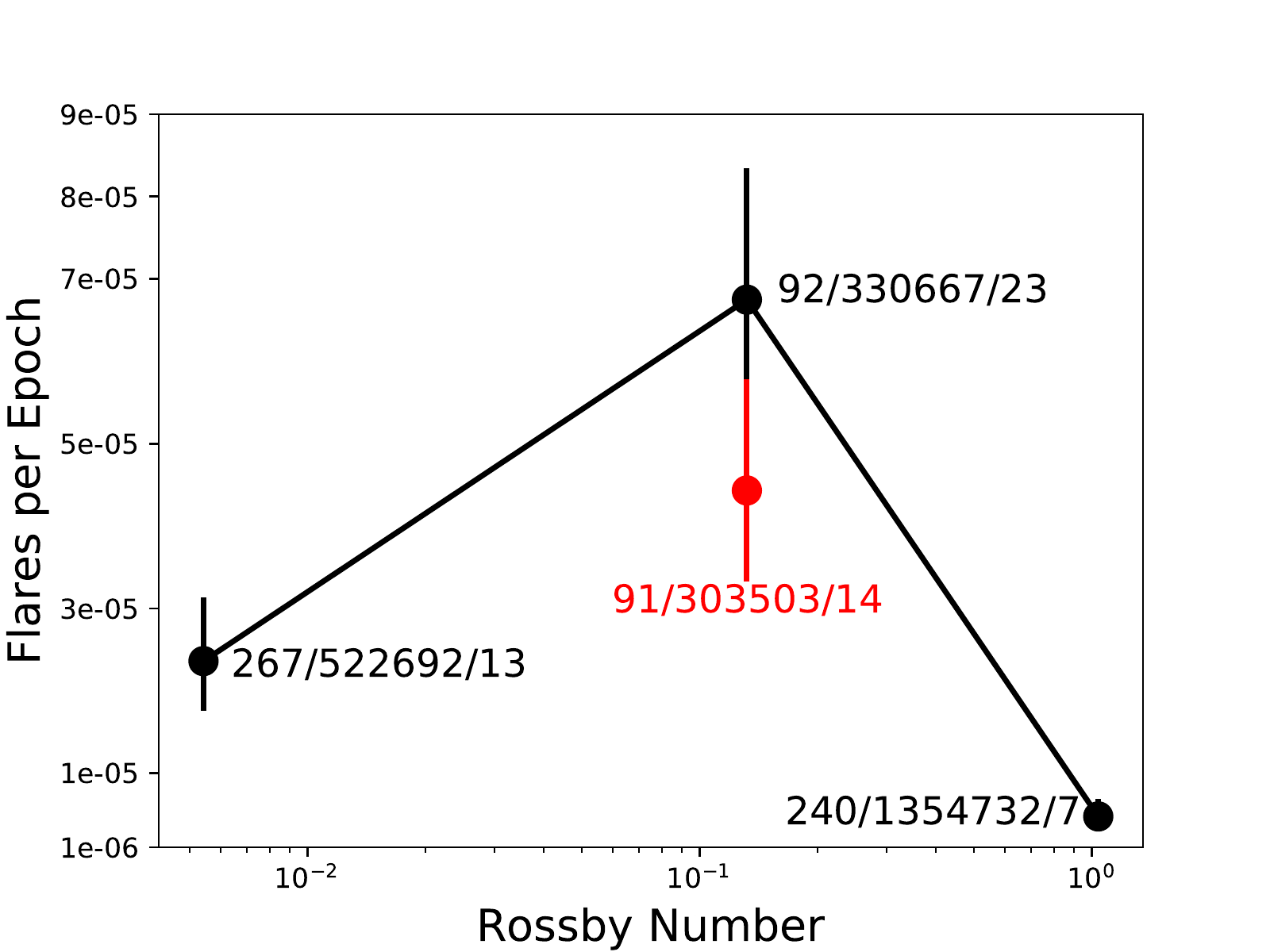}}%
	\hfill
	\subfloat[$p_\mathrm{Intermediate}-p_\mathrm{Fast}$]{\includegraphics[width=0.5\textwidth]{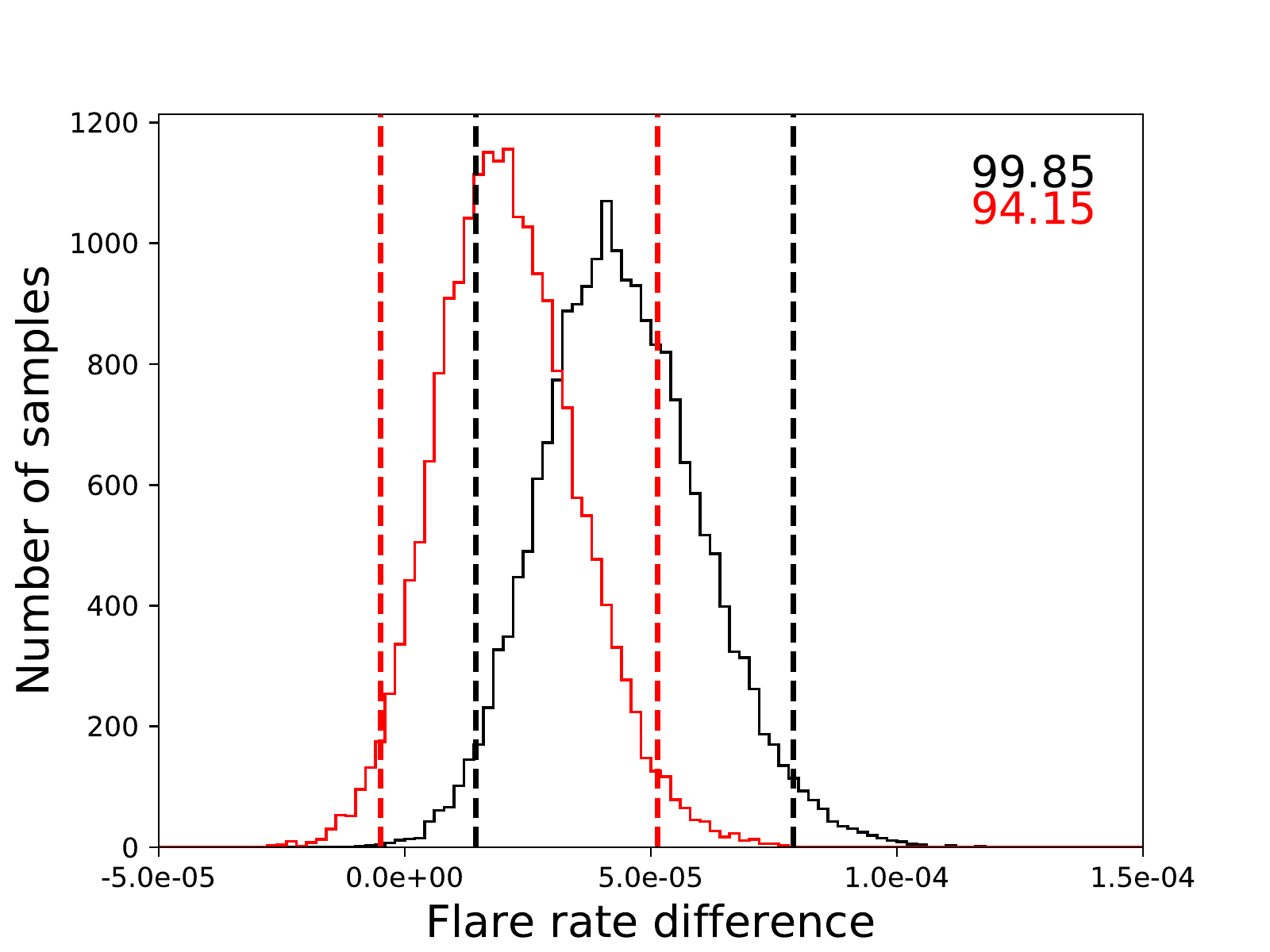}}%
	\\
	\subfloat[$p_\mathrm{Intermediate}-p_\mathrm{Slow}$]{\includegraphics[width=0.5\textwidth]{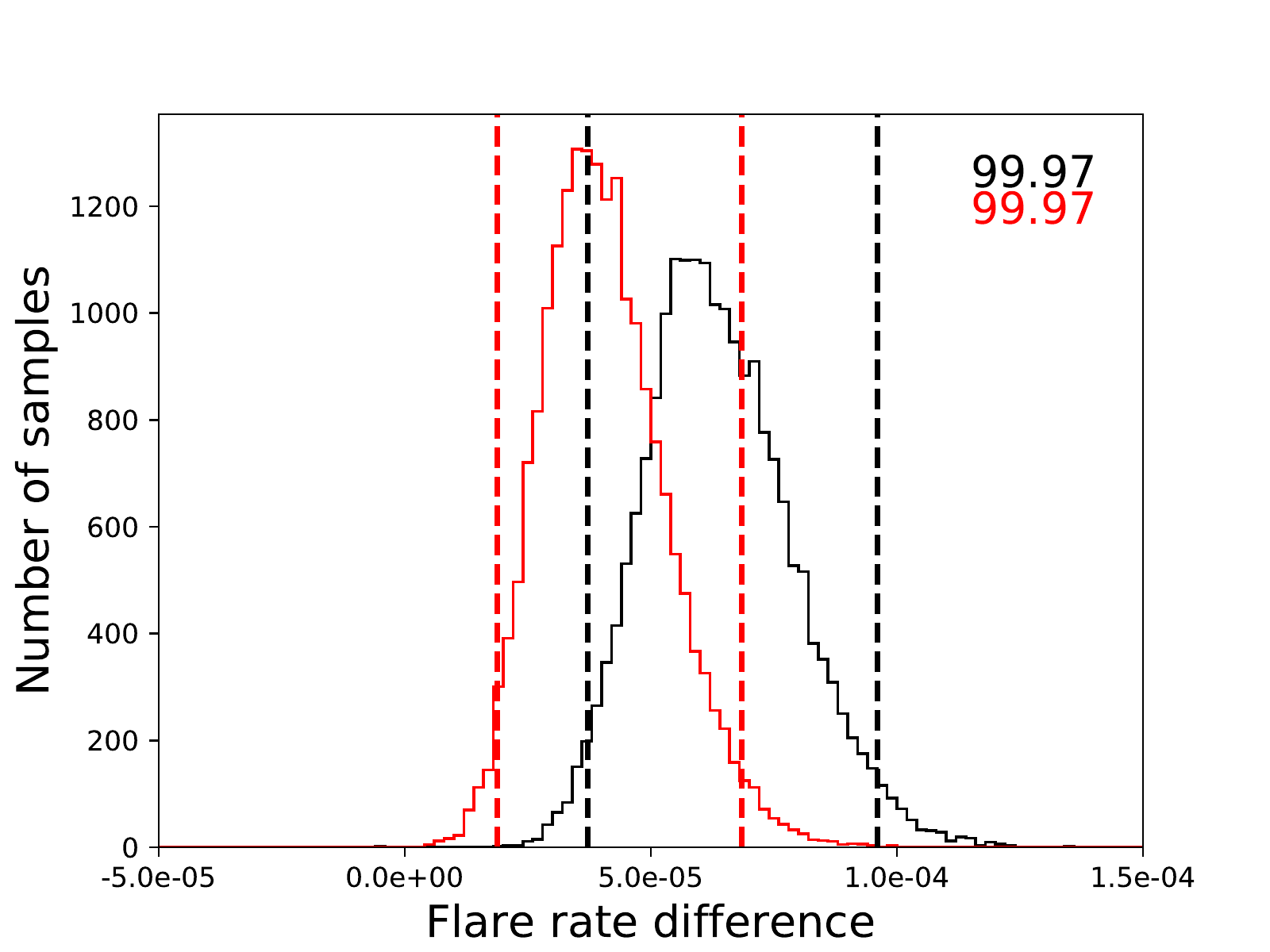}}%
	\hfill
	\subfloat[$p_\mathrm{Fast}-p_\mathrm{Slow}$]{\includegraphics[width=0.5\textwidth]{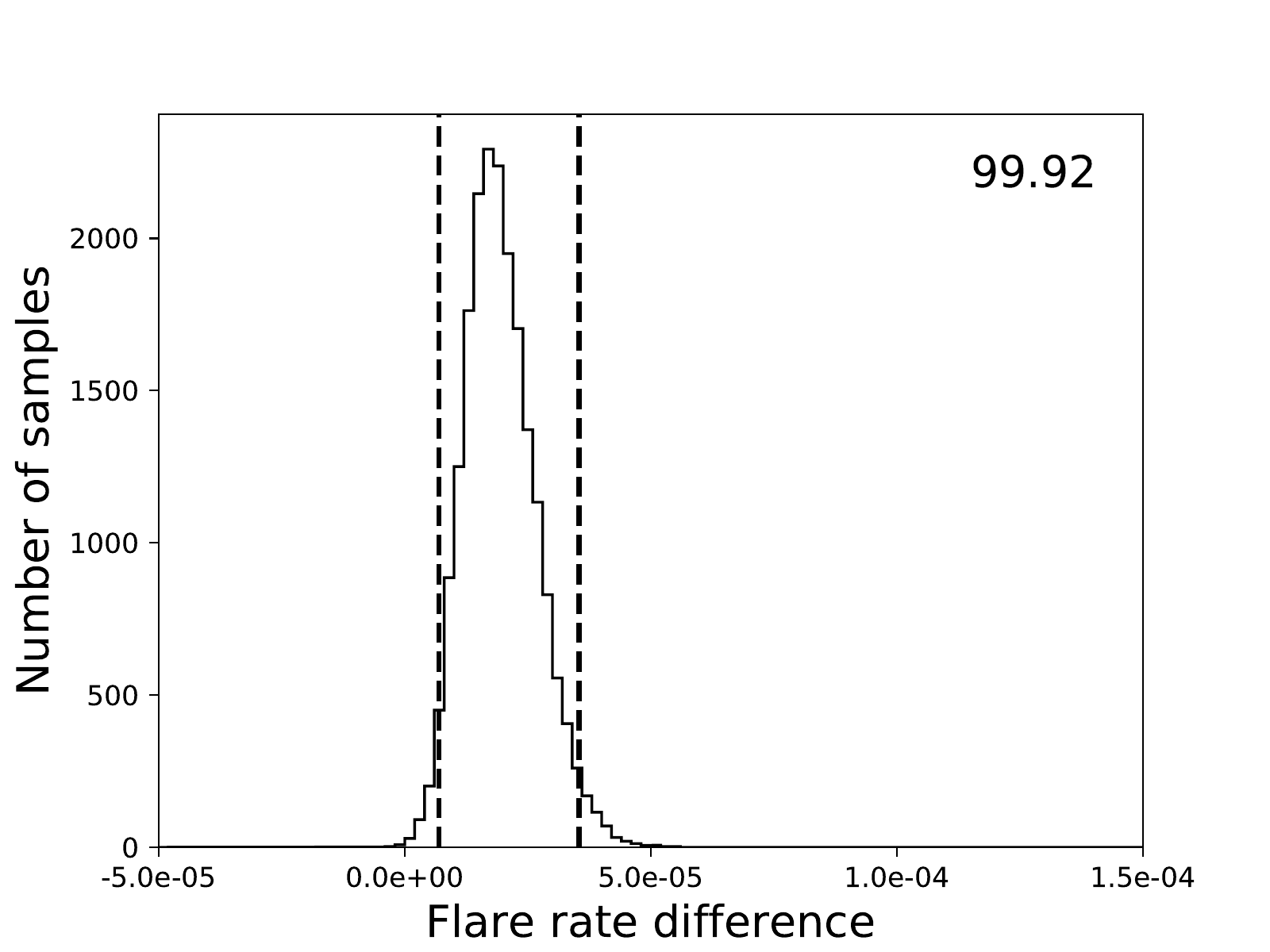}}%
	\caption{(a):The number of flares per observation as a function of $R_o$.  Periods are taken from Newton et al. (2018, submitted), \citetalias{Newton2016}, and references therein. The numbers in panel (a) show the numbers of stars/observations/flares in each bin.  The red point denotes the number of flares per epoch at the indicated $R_o$, but with flares from 2MASS J01374851+3823318 removed.  The errorbar in the last bin is smaller than the data point. (b), (c), and (d) show the distribution of the quantities $p_\mathrm{Intermediate}-p_\mathrm{Fast}$, $p_\mathrm{Intermediate}-p_\mathrm{Slow}$, and $p_\mathrm{Fast}-p_\mathrm{Slow}$ (see text for more details), calculated using the MCMC posteriors derived from (a).  The numbers indicate the size in percent of the one-sided credible region for which $p_i-p_j > 0$ in each panel.  Vertical dashed lines demarcate the 95\% credible region for their respective histograms.  The red histograms and numbers in (b) and (c) denote the same quantity, but with flares from 2MASS J01374851+3823318 removed.  }
	\label{fig:rothist}
\end{figure*}

\begin{figure*}
	\centering
	\subfloat{\includegraphics[height=0.2\textheight]{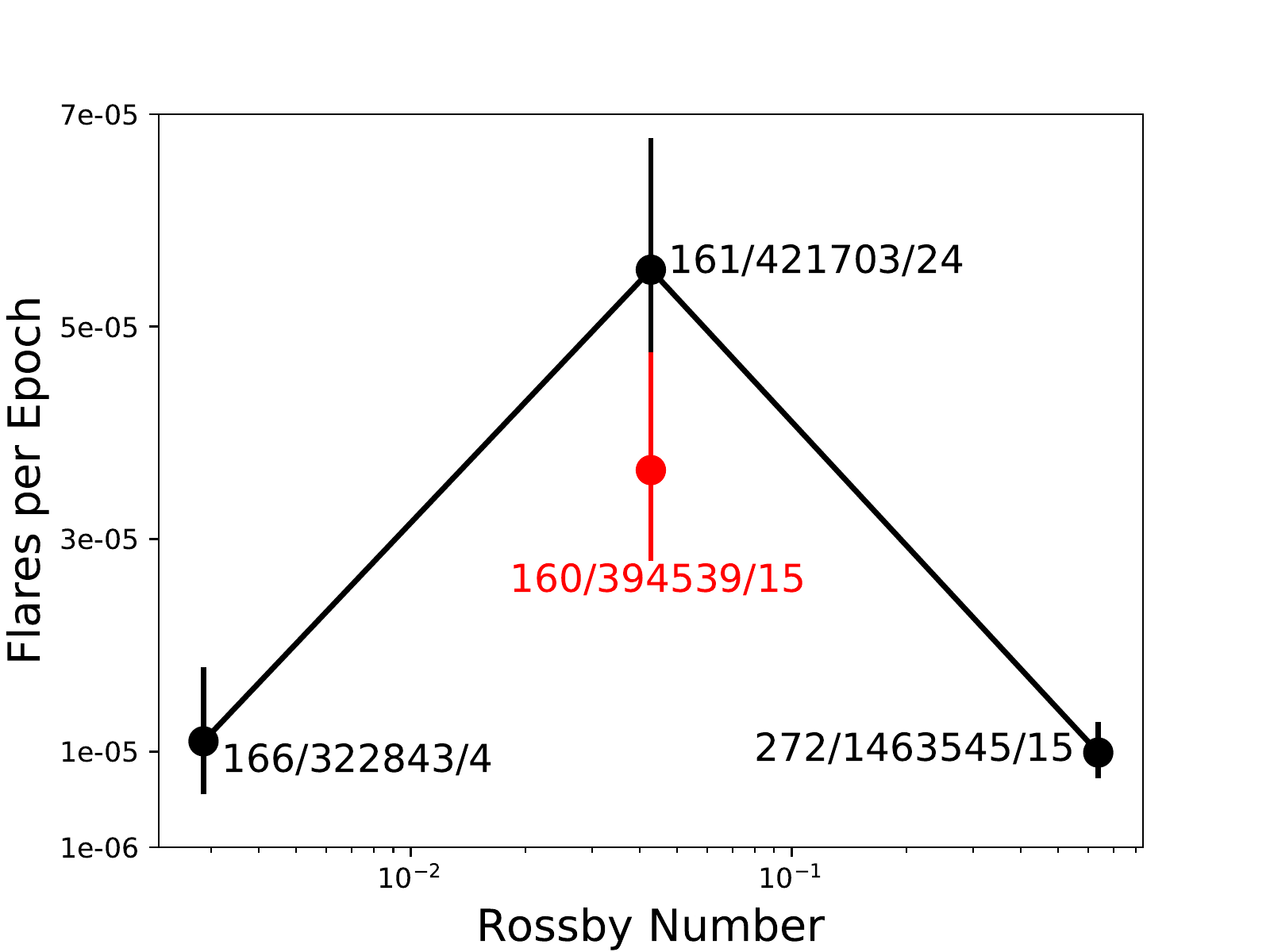}}%
	\subfloat{\includegraphics[height=0.2\textheight]{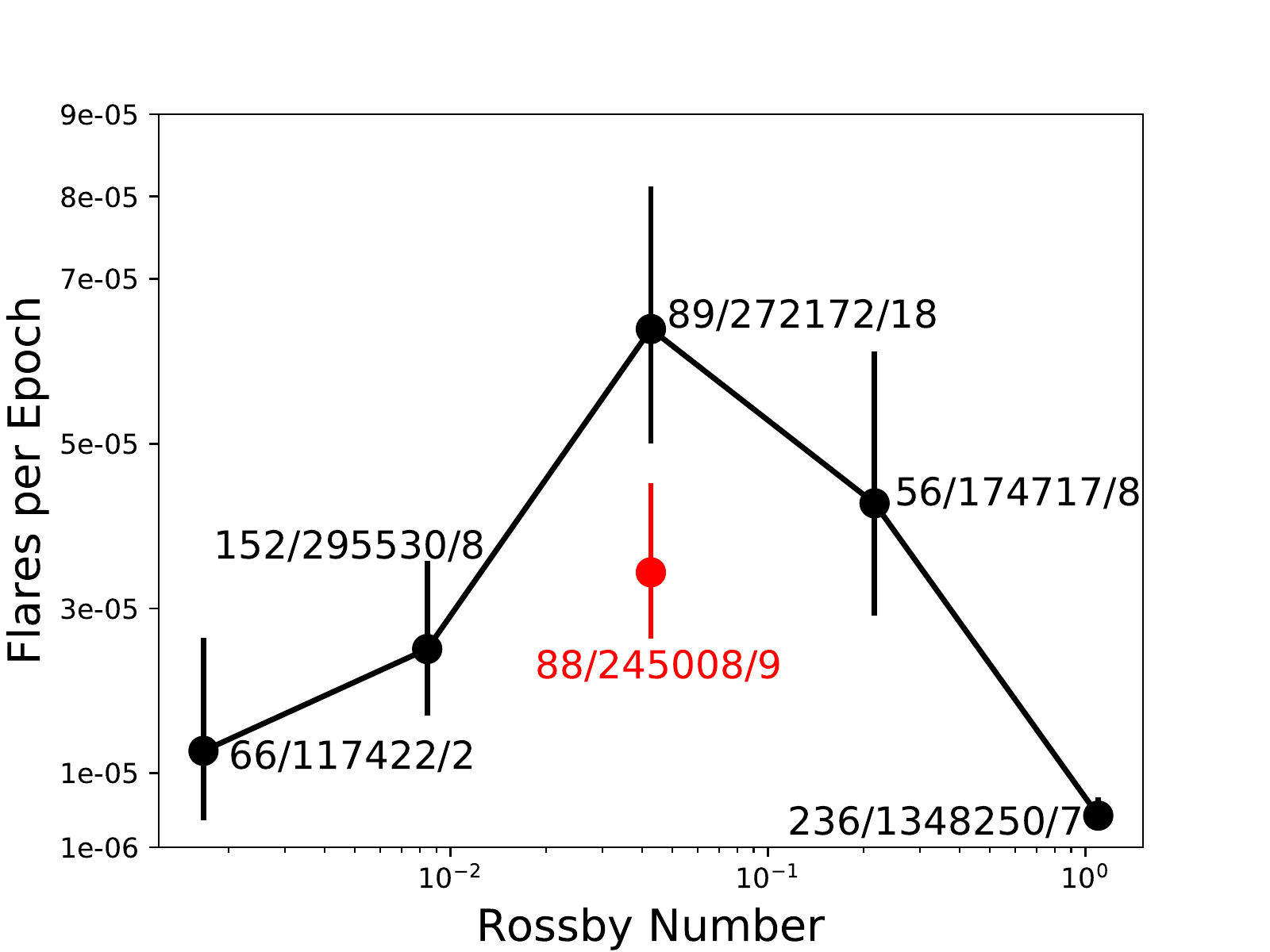}}%
	\\
	\subfloat{\includegraphics[height=0.2\textheight]{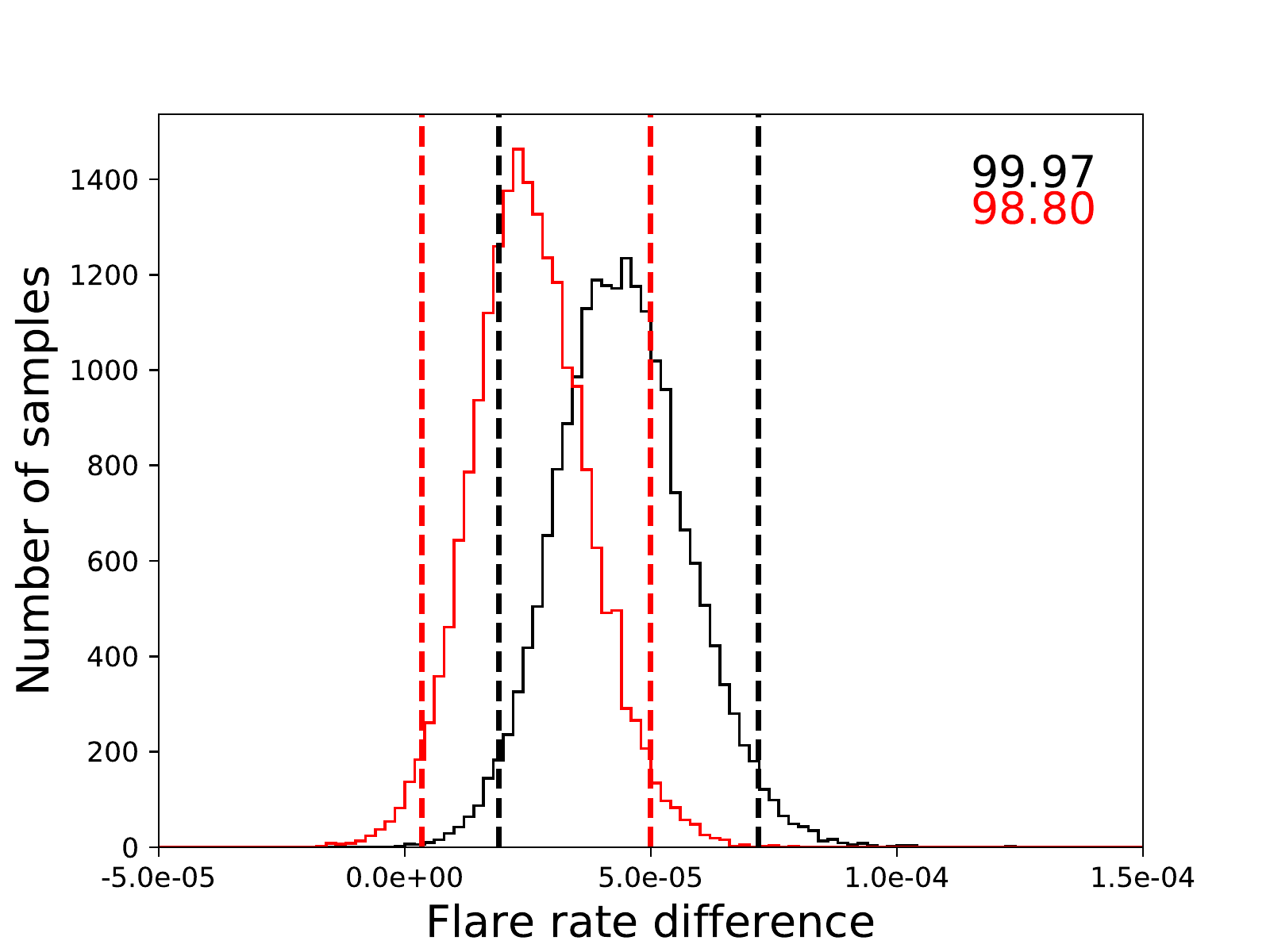}}%
	\subfloat{\includegraphics[height=0.2\textheight]{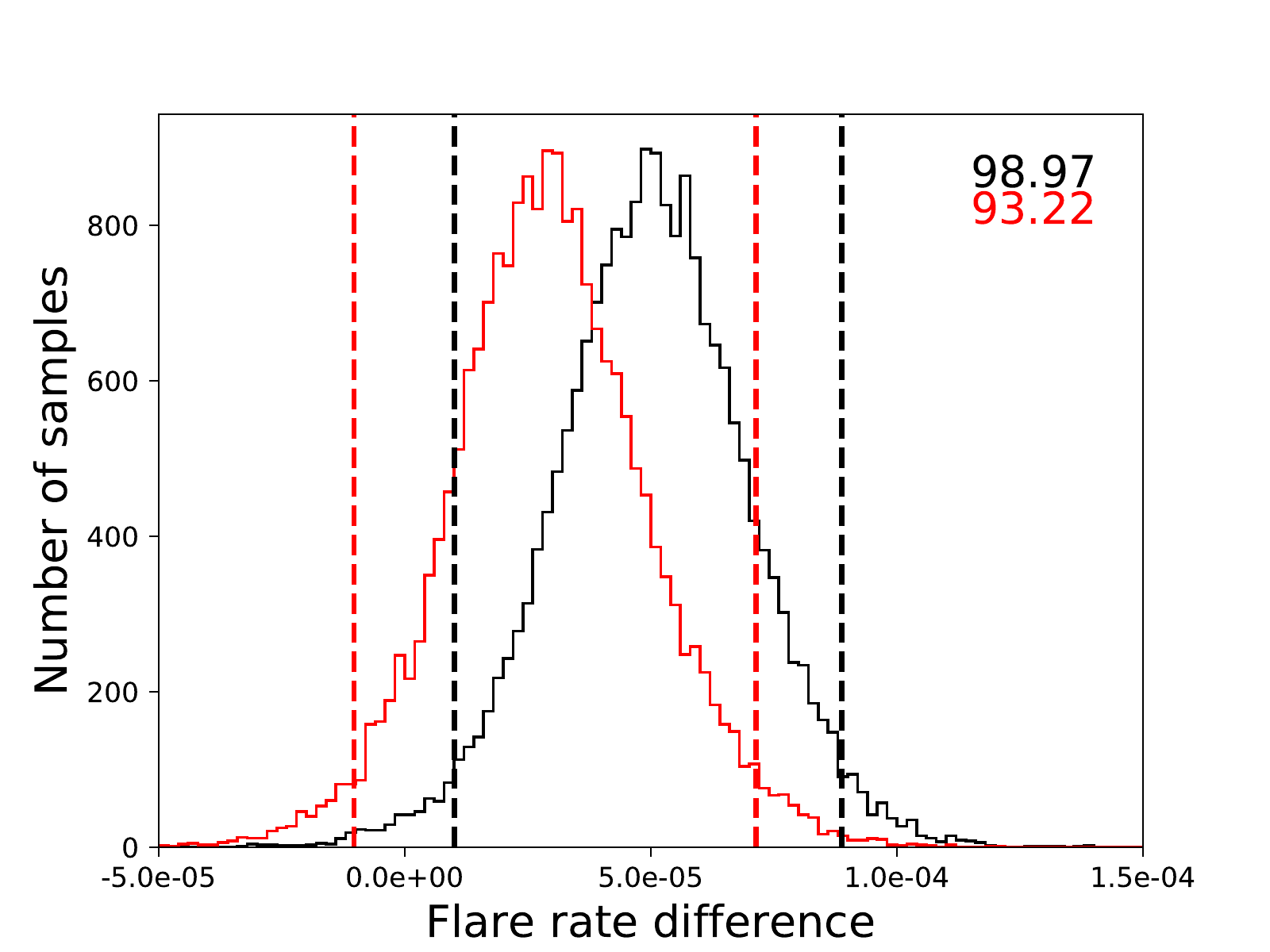}}%
	\\
	\subfloat{\includegraphics[height=0.2\textheight]{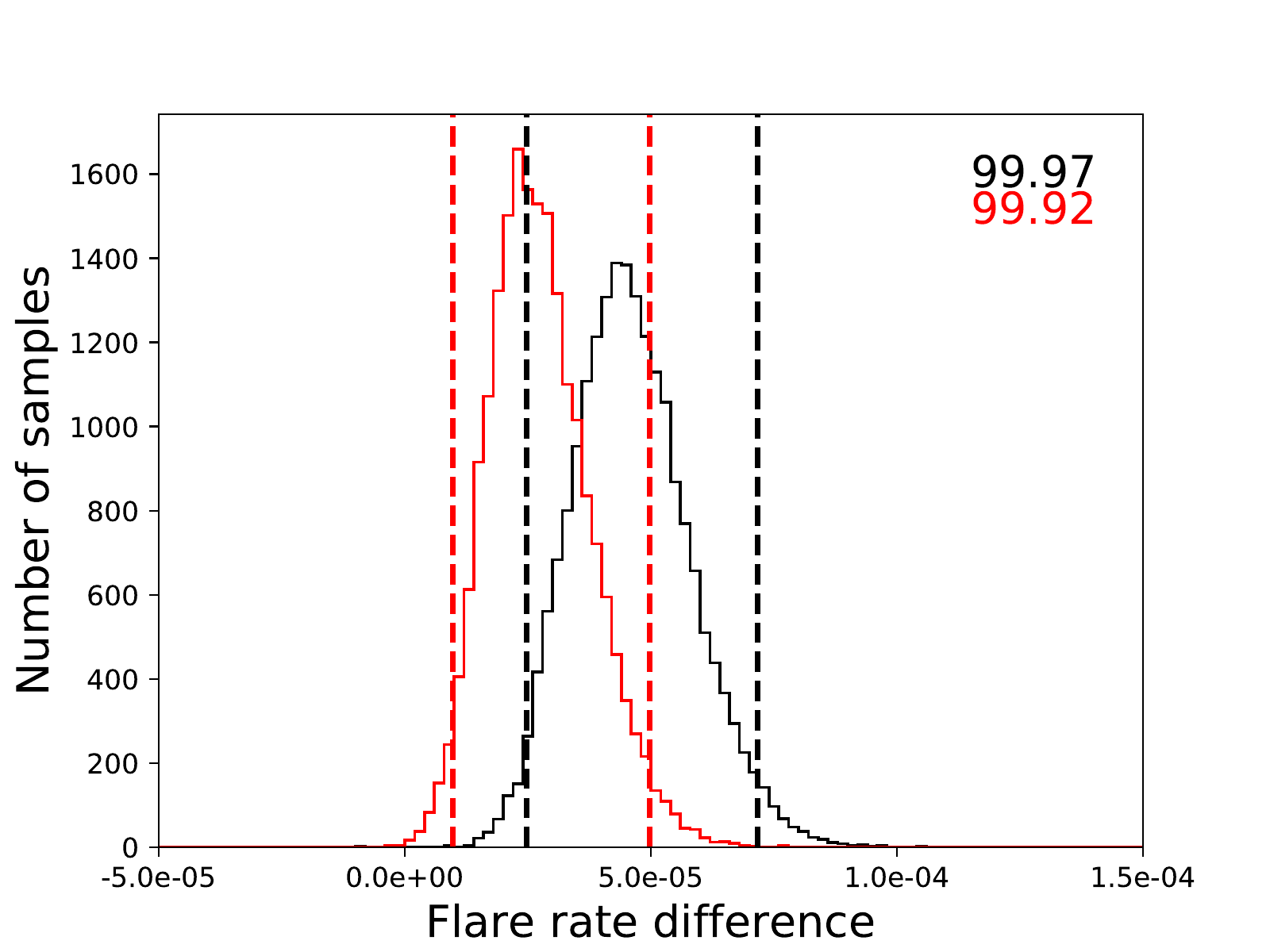}}%
	\subfloat{\includegraphics[height=0.2\textheight]{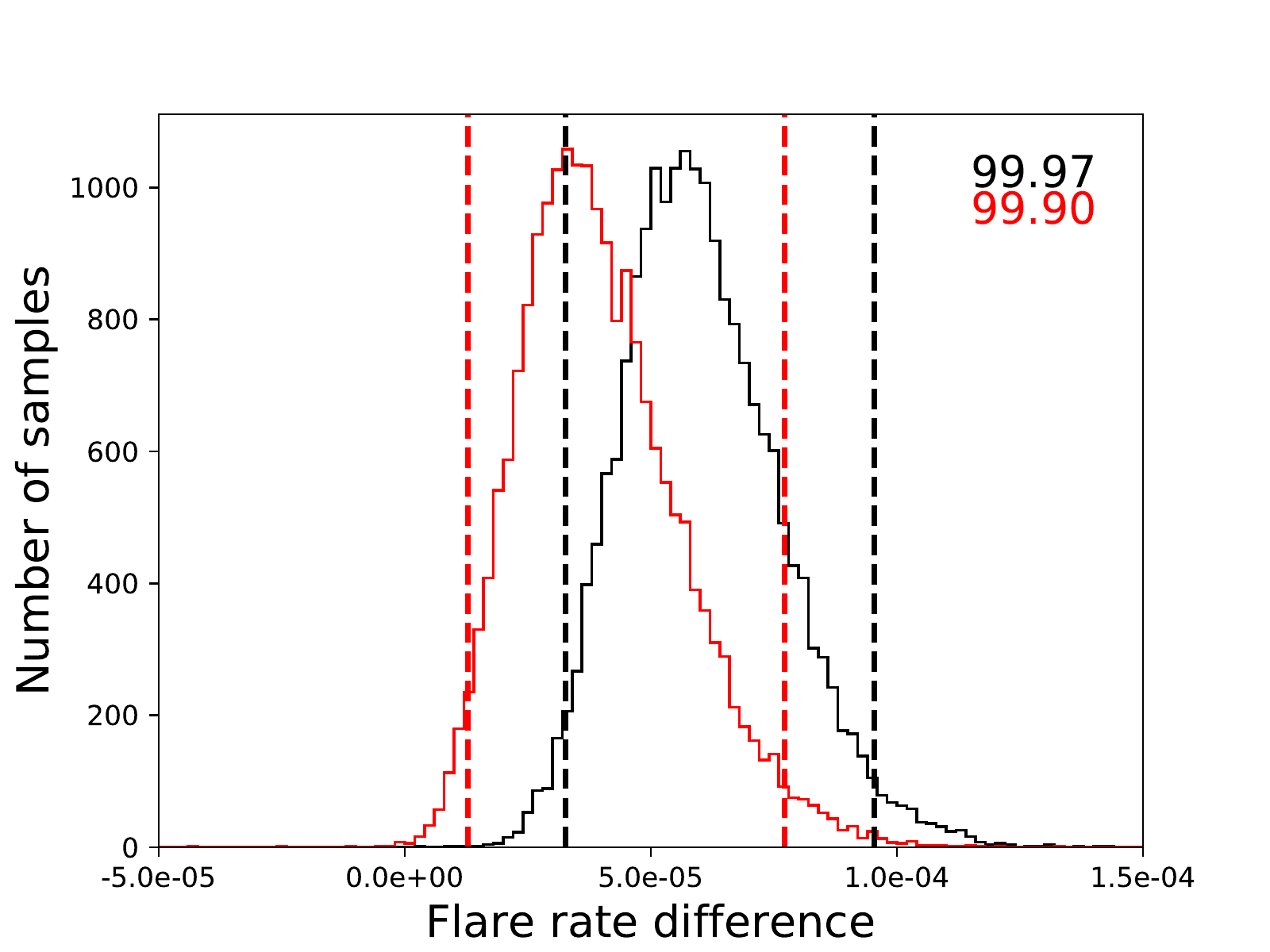}}%
	\\
	\setcounter{subfigure}{0}
	\subfloat[]{\includegraphics[height=0.2\textheight]{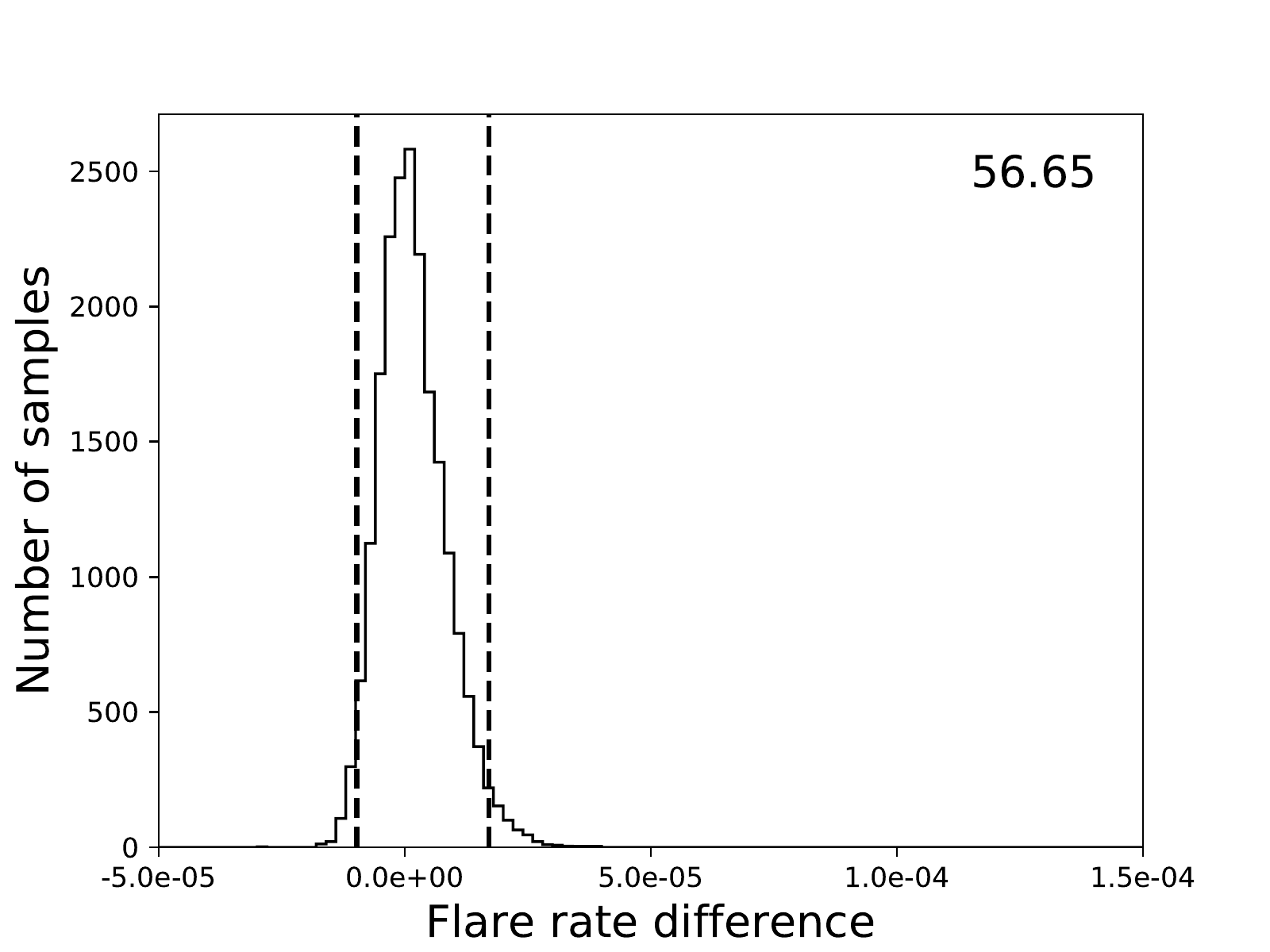}}%
	\subfloat[]{\includegraphics[height=0.2\textheight]{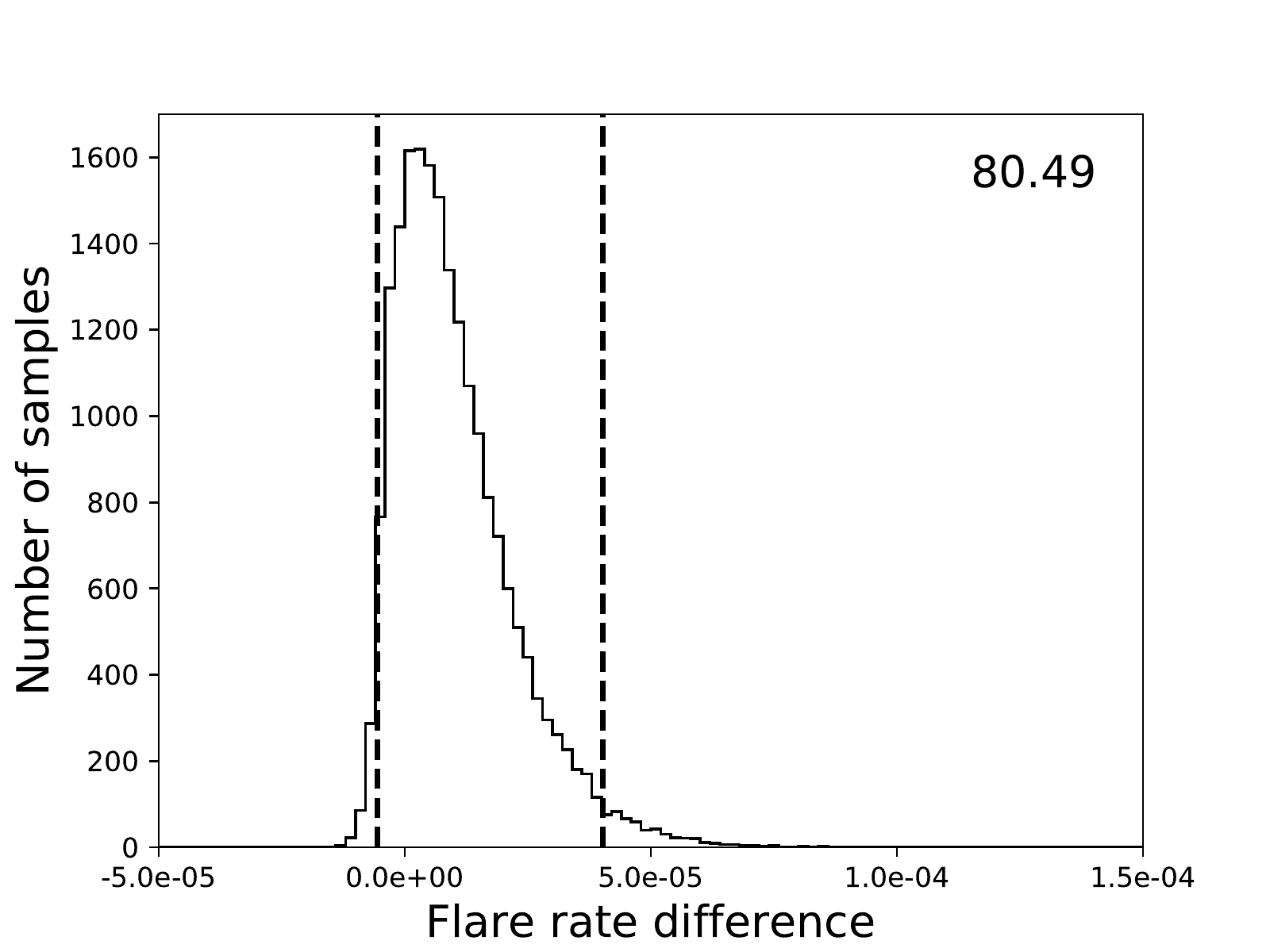}}%
	\\
	\caption{An exploration of the effect of binning on the results of Figure~\ref{fig:rothist}.  Stars are divided into 3 and 5 bins equally spaced in Log$_{10}(R_o)$ for columns (a) and (b), respectively.  From top to bottom, each figure in column (a) corresponds to subplot (a), (b), (c), and (d) from Figure~\ref{fig:rothist}.  For column (b), the bottom three figures represent $p_3-p_1$, $p_3-p_5$, and $p_1-p_5$ respectively, while the red histograms replace $p_3$ with p$_4$, where $p_i$ represents the estimate of the number of flares per epoch for bins 1-5, shown in the top row.  We emphasize that, as there is no longer any physical connection between bin locations and rotational classification, each bin can contain a polluted mixture of rotators.}
	\label{fig:rothistcompare}
\end{figure*}

To characterize differences in flare rates between stars as a function of $R_o$ and activity, we estimate the number of observed flares per observation marginalized across all stars within a bin of one of the aforementioned quantities. Uncertainties in Figure~\ref{fig:rothist}, \ref{fig:rothistcompare}, and Table~\ref{tab:act} are estimated using PyMC3 \citep{Salvatier2015}, assuming a binomial distribution with a flat prior in $\mathrm{ln}(p)$ from -15 to -2, where $p$ is the probability of detecting a flare in a given observation.  The number of successes in the binomial distribution is given by the number of flares observed, and the number of trials is the number of observations.  Errorbars in all figures denote the 16th and 84th quantiles for the results of the MCMC estimation, and the data points are plotted at the median of the posterior.

\begin{figure}[h!]
	\centering
	\includegraphics[width=\columnwidth]{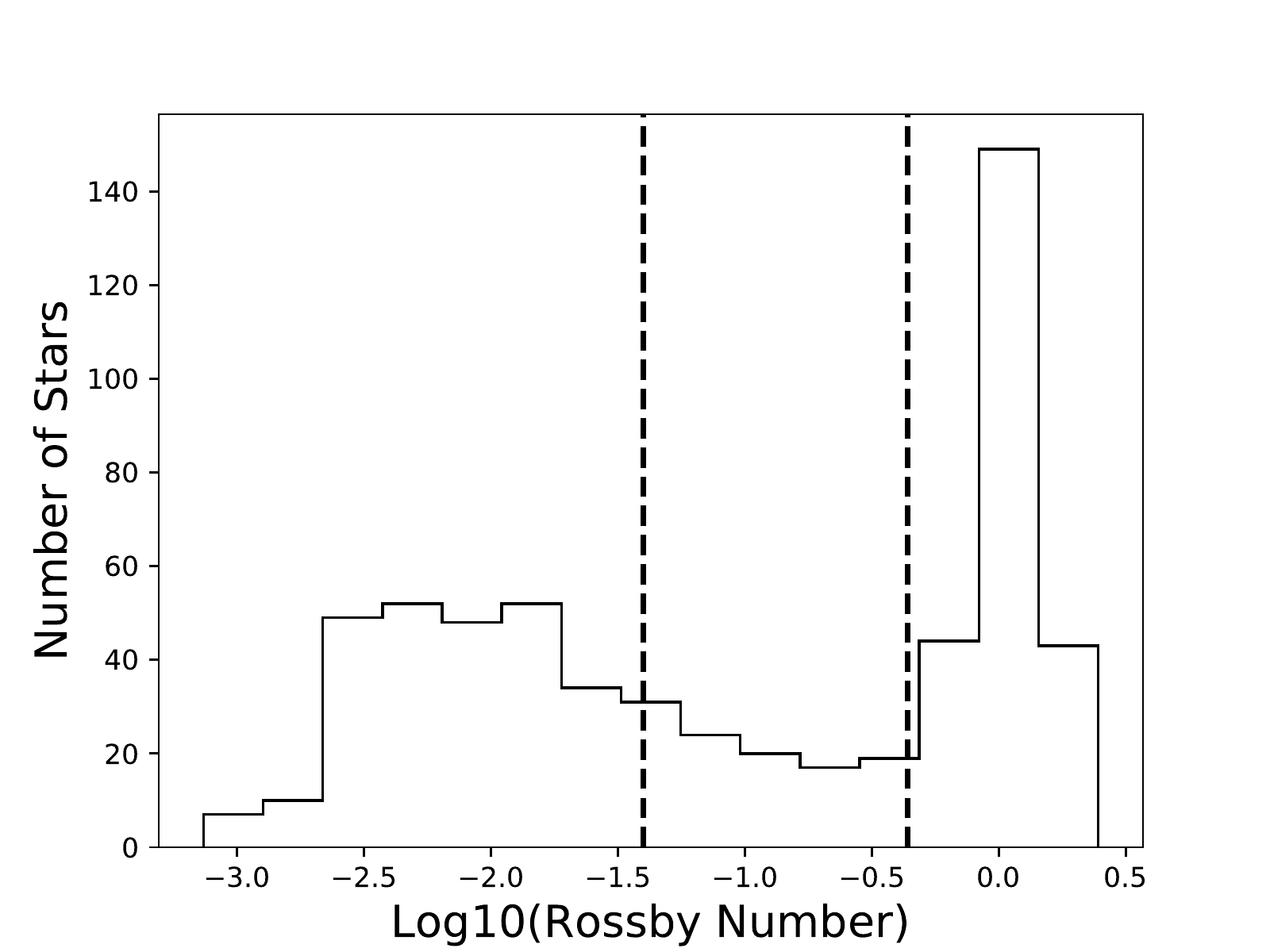}
	\caption{A histogram of the $R_o$ of stars used in this work.  We use the region between the vertical dashed lines as our population of intermediate rotators.}
	\label{fig:rossbyhist}
\end{figure}

Figure~\ref{fig:rossbyhist} shows the distribution of $R_o$ used in this work.  The vertical dashed lines separate the fast (leftmost region), intermediate (middle region), and slow (rightmost region) populations. Figure~\ref{fig:rothist} panel (a) shows the ratio of number of flares to number of observations binned by $\mathrm{log}_{10}(R_o)$.  The bins are defined at the black dashed lines in Figure~\ref{fig:rossbyhist}, given numerically as [0.04,0.44], {which approximate the breakpoint between the various rotational groups}.  The convective overturn timescale ($\tau_\mathrm{conv}$) needed to calculate $R_o$ is estimated using equation 11 of \citet{Wright2011}, which is valid for $0.09<M/M_\odot<1.36$.  Masses for calculating $\tau_\mathrm{conv}$ are estimated from absolute K magnitudes, using the relation of \citet{Benedict2016}.  Of particular note is the rise in flare rate at intermediate $R_o$, the regime in which a rapid spindown mechanism may be acting.

We use the results of the MCMC estimator from Figure~\ref{fig:rothist} to sample the distribution of the function $p_\mathrm{Intermediate}-p_\mathrm{Fast}$ in order to determine to what extent the data support an increase in flare rates from low $R_o$ (fast rotators, the first bin in each panel of Figure~\ref{fig:rothist}) to intermediate $R_o$ (the central bin in Figure~\ref{fig:rothist}).  The distribution of $p_\mathrm{Intermediate}-p_\mathrm{Fast}$ is shown in panel (b) of Figure~\ref{fig:rothist}.  The black vertical lines denote the 2.5\% and 97.5\% quantiles, which we use to represent the 95\% credible region.  The numbers on the figure represent the size of the one-sided credible region (in percent) in which the posterior is greater than zero.  The red histogram, numbers and vertical lines denote the same quantities, except that the contribution from 2MASS J01374851+3823318 has been removed.  The largest one-sided credible region that does not include zero is the 99.85\% credible region -- we can therefore state that the flare rate of intermediate rotators, as observed by MEarth, is greater than that of fast rotators at the 99.85\% confidence level.  When the contributions from 2MASS J01374851+3823318 are neglected, this number falls to 94.15\%.  We stress that there is no \textit{a priori} reason to remove 2MASS J01374851+3823318 from the intermediate rotator population. Additionally, high $R_o$ stars show almost no signs of flaring, with the highest $R_o$ rotator (2MASS J17462934-0842362, known SB2) showing a flare having a period of 66 days, and a mass of 0.31$M_\odot$ ($R_o=1.2$), though its EW of H-$\alpha$ is -1.16, which we classify as an active star. The longest period rotator showing a flare is Proxima Centauri (2MASS J14294291-6240465, $0.12M_\odot$, \prot=$88.5$ days, $R_o=0.75$, active in H-$\alpha$), which is still in the process of spinning down, due to its low mass.

We also examine the distributions of $p_\mathrm{Intermediate}-p_\mathrm{Slow}$ (panel (c), Figure~\ref{fig:rothist}) and $p_\mathrm{Fast}-p_\mathrm{Slow}$ (panel (d), Figure~\ref{fig:rothist}) in a similar manner.  Unsurprisingly, the distribution shifts to the right for $p_\mathrm{Intermediate}-p_\mathrm{Slow}$, as slow rotators flare much less frequently than rapid or intermediate rotators. The distribution of $p_\mathrm{Fast}-p_\mathrm{Slow}$ also shows the expected behavior, in that there is a high probability that $p_\mathrm{Fast}>p_\mathrm{Slow}$.

When using histograms, there is always a danger that any conclusions drawn from the data may be an artefact induced by one's choice of binning.  To examine whether this result is a strong function of bin choice, Figure~\ref{fig:rothistcompare} repeats the analysis from Figure~\ref{fig:rothist}, except that the stars are divided into 3 (column (a)) or 5 (column (b)) linearly spaced bins.  We expect that such a binning strategy might alter our results slightly, as the group of stars we are classifying as intermediate in Figure~\ref{fig:rothistcompare} are in actuality comprised of a mixture of rotator classes.  In this case, the difference between the first and last bins of each plot (bottom row) becomes less distinct, though the probability that the middle bins differ from the extremes remains large.  Note that in column (b), the red histogram that neglects flares from 2MASS J01374851+3823318 is constructed using $p_4$ instead of $p_3$, where $p_i$ represents the number of flares per epoch for bin $i$.  This is because both $p_3$ and $p_4$ contain a population of intermediate rotators, and when excluding 2MASS J01374851+3823318 from $p_3$, flares are probable in bin four.

\begin{figure*}[h!]
	\includegraphics[width=0.9\textwidth]{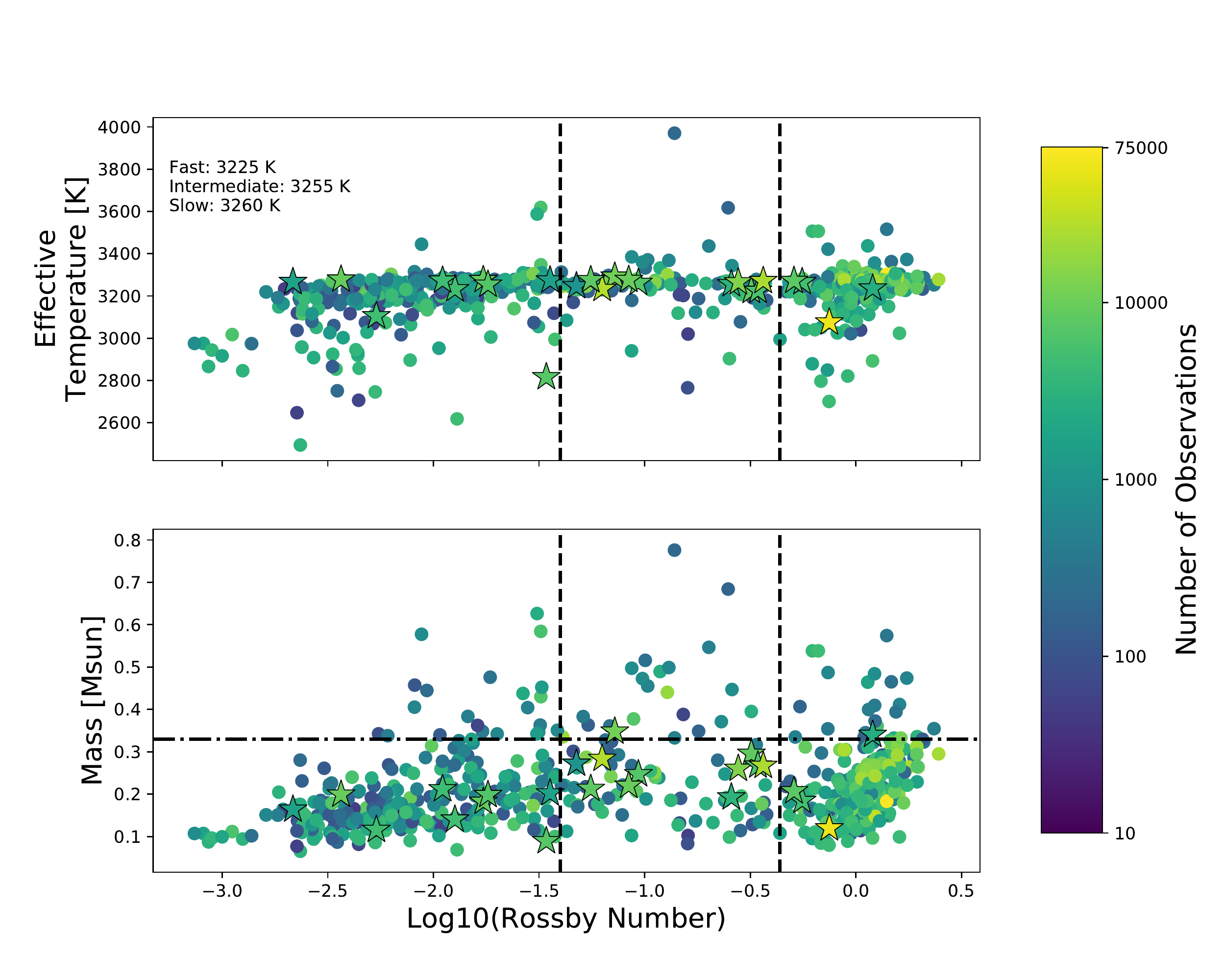}%
	\caption{Mass and effective temperatures for stars in the MEarth sample. Points are color coded by the number of observations for that star.  Stars showing one or more flares are denoted as black-outlined stars in the plot.  The vertical dashed lines in both panels denote the cutoff values of $R_o$ for the fast, intermediate, and slow groups.  The horizontal dot-dashed line in the lower panel shows the approximate location of the fully-convective limit (0.33 $M_\odot$).}
	\label{fig:mass-teff-rossby}
\end{figure*}

\begin{figure}[h!]
	\includegraphics[width=\columnwidth]{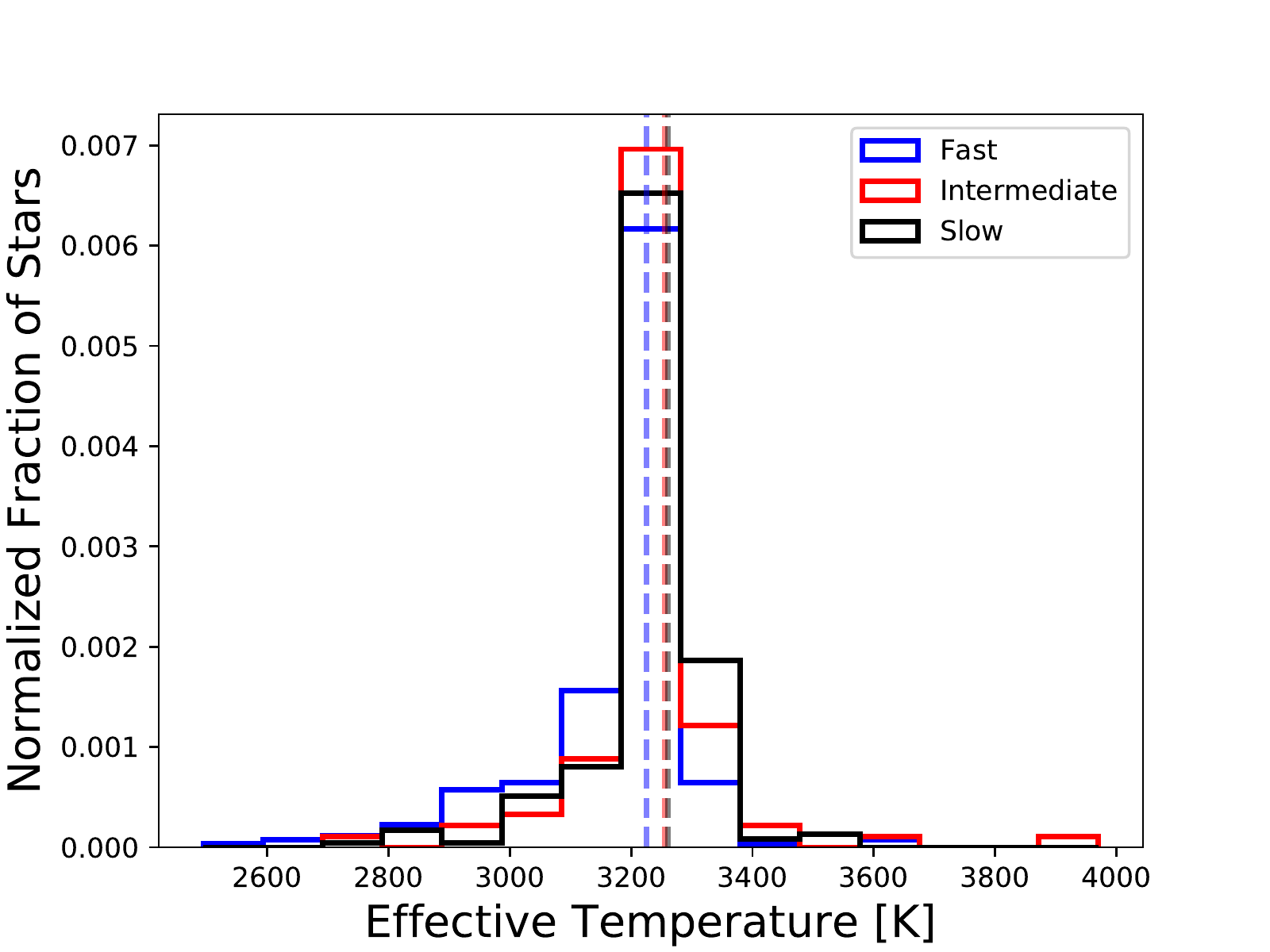}
	\caption{Effective temperature distributions of the fast (blue), intermediate (red), and slow (black) populations.  Vertical colored dashed lines show the median effective temperatures for their respective histograms.  The median temperatures for each histogram are 3225 K (fast), 3255 K (intermediate), and 3260 K (slow).}
	\label{fig:teff-rossby-dists}
\end{figure}

{
\subsection{Investigating Sample Biases}
In addition to arising as an artifact due to histogram bin choice, we also investigate the possibility that the differences in flare rates between the 3 classes of rotators arise from differing physical parameters for the groups.  In particular, Figure~\ref{fig:mass-teff-rossby} shows the effective temperatures and masses of the rotators examined in this work.  As the top panel shows, most stars in MEarth occupy a relatively narrow band of temperatures around 3200 K.  Figure~\ref{fig:teff-rossby-dists} shows the distribution of effective temperatures for each of the rotator populations.  The median temperatures for each group are 3225 K (fast), 3255 K (intermediate), and 3260 K (slow).  While it is known that cooler stars flare more frequently \citep{Kowalski2009}, the narrow spread in effective temperatures does not provide an explanation for the observed difference in flare rates.}

\subsection{Activity and Flaring}
Table~\ref{tab:act} shows the measured number of flares per epoch for both active and inactive populations.  Of the 34 flaring stars, {26 have EW measurements}. Active stars are defined as having an H-$\alpha$ EW less than $-1$ (where negative EW implies emission), while inactive stars have an EW greater than $-1$.  H-$\alpha$ EW measurements are taken from \citet{Newton2017} and sources listed therein.  We find {44 flares on active stars}, and 1 flare on an inactive star (2MASS J15012274+8202252, $0.26M_\odot$, no rotation period).  However, there is some evidence to suggest that this star is a binary (Winters et al., 2018, submitted).  If it is in fact a binary system, then the presence of a companion could be inducing flares that we observe on the primary, which are then observed by MEarth.  In this case, we would observe no flares on inactive stars.  MEarth is approximately {75x more likely} to find a flare in a given observation on an active star than on an inactive star.

\begin{deluxetable}{cl}
\tablecaption{Active and Inactive Flares per Epoch\label{tab:act}}
\tablehead{\colhead{State} & \colhead{Flares per epoch}}
	\startdata
Active & $(4.77 \, ^{+0.324}_{-0.355})\times10^{-5}$ \\
Inactive & $(6.36 \, ^{+1.91}_{-2.81})\times10^{-7}$ \\
	\enddata	
\end{deluxetable}

\section{Discussion and Conclusions}
\label{sec:conclusion}
\subsection{Connections with Kepler Studies}
\label{sec:connection}

{A number of studies using data from the \textit{Kepler} mission have considered the connection between flares and stellar rotation. \citet{Davenport2016} searched all available \textit{Kepler} light curves for flares, with the most directly relevant results being those comparing $L_\mathrm{flare}/L_{Kp}$ (the ratio of flare luminosity to \textit{Kepler} luminosity) to $R_o$.  The author observes a behavior similar to that of $L_{\mathrm{H}\alpha}/L_\mathrm{Bol}$ vs $R_o$, with stars at low $R_o$ showing saturated values, and stars at high $R_o$ showing decreased relative flare luminosity.  This is qualitatively not in conflict with our result from Figure~\ref{fig:rothist} because decreased flare luminosity, which is calculated (in part) by summing the equivalent durations of the observed \textit{Kepler} flares, can be caused both by decreased frequency and by decreased energy emission from flares.  This saturation behavior was not observed in the reddest bin tested ($2.5<g-i<3.0$, M2-M4 spectral types), as their sample did not contain enough short period rotators to observe the saturation region.  The \textit{Kepler} M dwarfs they studied are also mostly more massive than the fully convective limit, which could have an impact on the dynamo mechanism and flare generation processes compared to our sample.  \citet{Candelaresi2014} previously found a peak at $R_o \simeq 0.1$ in superflare ($E_\mathrm{Flare} > 5\times10^{34}$ erg) rates in G-, K-, and M-dwarfs from the \textit{Kepler} dataset.  The \citet{Candelaresi2014} result notably does not show the saturation region as in \citet{Davenport2016}, and consists instead of only rising and falling regions.  \citet{VanDoorsselaere2017} also extract and examine flares from \textit{Kepler} data, and show in their Table 4 a comparison of incidence rate vs Log$(R_o)$.  They also tenatively find evidence for a peak at a Log$(R_o) \simeq -1.2$.  This location is comparable to our intermediate peak at Log$(R_o) \simeq -0.9$ and also to that found by \citet{Candelaresi2014}.  However, as in the case of \citet{Davenport2016}, their sample includes relatively few stars with effective temperatures less than approximately 3500 K, meaning that there is little overlap with the sample of stars used in this work.  \citet{Yang2017} examine the dependence of $L_\mathrm{flare}/L_\mathrm{bol}$ on rotation period in \textit{Kepler} stars, finding that the ratio has an inflection point at a period of about 4 days.  The sample in \citet{Yang2017} is subject to the same selection bias as the other \textit{Kepler} studies, in that they have few stars below an effective temperature of approximately 3200 K, while our sample is composed almost entirely of stars in this temperature range.}
	
{The decrease in flare activity for faster rotators at $R_o \simeq 0.1$ is similar to an effect seen in coronal X-ray emission known as the supersaturation phenomenon, as demonstrated in $L_\mathrm{X}/L_\mathrm{bol}$ vs. \prot\ diagrams.   \citet{Wright2011} argue against X-ray supersaturation being exhibited in M dwarfs, noting that to be in this regime requires rotation periods approaching break-up speeds, and faster than the observed periods.  A study by \citet{Jeffries2011} examined X-ray emission in rapidly rotating M dwarfs, and found no evidence for supersaturation for Log$(R_o) > -2.3$.  There is also no guarantee that the physical mechanism responsible for supersaturation in coronal X-rays is the same as the one responsible for the drop in flare rates for short period rotators.}

\subsection{Summary}
\label{sec:summary}
There are two conclusions to be drawn from the observed relation between rotation and flaring.  First, there appears to be a rise in flare rates at intermediate $R_o$.  Second, the slowest rotators produce large flares infrequently.  The change in observed flare rates between fast, intermediate, and slow rotators could be due to a few effects.  If it is the case that mid-to-late M dwarfs enter a state of rapid spindown at some stage in their lives, the physical process governing spindown could also trigger flares.  Flares on intermediate rotators could also occupy a different region of phase space (longer lasting, higher amplitudes) than flares on fast rotators, leading to a higher detection efficiency in MEarth for these flares. A combination of these effects would also explain our results.

The presence of this rise in detected flare rate at $R_o\simeq0.1$ also coincides approximately with the minimum in spin-down time reported by \citet{See2017}.  An important note is that \citet{See2017} contains only 8 stars below 0.33 $M_\odot$, so the degree of overlap in terms of dynamo mechanisms between their sample and the one presented here is not clear.  In addition, the stars in \citet{See2017} are all within the rapid rotating population as shown in \citet{Newton2017}.

The growing number of studies showing bimodal distributions of surface magnetic field topologies may shed some light on the results presented here.  We suggest that as the star spins down, changing field topology may manifest itself as increased flaring in the star. The period of increased flaring ends when the field settles in a much weaker state, thereby suppressing flares.

The key to understanding the angular momentum evolution of mid-to-late M dwarfs from young (roughly 2 Gyr) and rapidly rotating to old (roughly 5 Gyr) and slowly rotating is the evolution of the intermediate rotators. In particular, measurements of the total magnetic field strength, as well as the topology (dipolar vs multipolar) of these intermediate rotators will yield valuable insights to the evolution of these stars.

It may be possible to use observables such as flare rate and rotation period to determine the characteristics of the mechanism that determines magnetic field topology in these stars.  In particular, if the dynamo powering late M dwarfs indeed has bistable regimes (as proposed in \citealt{Gastine2013}) determined by initial conditions, then one might expect to find two populations in the period distribution of stars with a given age: the first being relatively slowly rotating, due to having resided on the dipolar branch of the dynamo, and the second being more rapidly rotating, due to residing on the multipolar branch for their lifetime.  However, if an oscillatory mechanism is responsible, as in \citet{Kitchatinov2014}, then (assuming the timescale for oscillation is much shorter than the stellar spindown timescale), the star should lose angular momentum at a roughly constant rate over long periods of time.  This would manifest itself as a single population of rotators at a given age, rather than a bimodal one.  The flaring properties of these stars would likely also be different, in the case that flaring is exacerbated by the spindown mechanism.  In the future, the Large Synoptic Survey Telescope (LSST, \citealt{Ivezic2008a}), will find low mass rotators, which may help to differentiate between the two models.  The Transiting Exoplanet Survey Satellite (TESS, \citealt{Ricker2014}) will also be a bountiful source of both rotation periods and flares with which to expand this work. Additionally, because TESS targets will in general be closer and brighter than those observed by LSST, they will be more amenable to follow-up from ground- and space-based telescopes.

\section{Acknowledgments}
The MEarth Team gratefully acknowledges funding from the David and Lucille Packard Fellowship for Science and Engineering (awarded to D.C.). This material is based upon work supported by the National Science Foundation under grants AST-0807690, AST-1109468, AST-1004488 (Alan T. Waterman Award), and AST-1616624. This publication was made possible through the support of a grant from the John Templeton Foundation. The opinions expressed in this publication are those of the authors and do not necessarily reflect the views of the John Templeton Foundation.  N.M. is supported by the National Science Foundation Graduate Research Fellowship Program, under NSF grant number DGE1745303. N.M. thanks the LSSTC Data Science Fellowship Program; his time as a Fellow has greatly benefited this work. E.R.N. is supported by an NSF Astronomy and Astrophysics Postdoctoral Fellowship under award AST-1602597.  We thank the anonymous referee for their suggestions that substantially improved this manuscript.

\bibliography{mybib}

\end{document}